\documentclass[twocolumn,dvipsnames]{aastex62}
\usepackage{amsmath,amstext}
\usepackage{apjfonts}
\usepackage[figure,figure*]{hypcap}
\usepackage{ulem}
\usepackage{xspace}
\usepackage{lineno}
\usepackage{xfrac}
\usepackage{graphicx}
\usepackage{natbib}
\usepackage{multimedia}
\usepackage{amssymb}
\usepackage{amsmath}
\usepackage{array,booktabs}
\usepackage{mathptmx}
\usepackage{booktabs}
\usepackage{hyperref}
\usepackage{float}
\usepackage{todonotes}
\usepackage{multirow}
\newcommand{\mbh}{\,{M_{\rm BH}}}
\newcommand{\abh}{\,{a_{\rm BH}}}
\newcommand{\erg}{\,{{\rm erg}}}
\newcommand{\msun}{\,{M_{\odot}}}
\newcommand{\s}{\,{{\rm s}}}

\newcommand{\km}{\,{{\rm km}}}
\newcommand{\ms}{\,{{\rm ms}}}
\newcommand{\egw}{\,{E_{\rm GW}}}
\newcommand{\fgw}{\,{f_{\rm GW}}}
\newcommand{\tacc}{\,{t_{\rm acc}}}

\shorttitle{vigorous coherent gravitational waves from cooled collapsar cisks}
\shortauthors{Gottlieb, Levinson \& Levin}

\begin{document}

\title{In LIGO's Sight? Vigorous Coherent Gravitational Waves from Cooled Collapsar Disks}

    \author[0000-0003-3115-2456]{Ore Gottlieb}
	\email{ogottlieb@flatironinstitute.org}
    \affil{Center for Computational Astrophysics, Flatiron Institute, 162 5th Avenue, 6th floor, New York, NY 10010}
    \affil{Physics Department and Columbia Astrophysics Laboratory, Columbia University, 538 West 120th Street, New York, NY 10027}

    \author[0000-0001-7572-4060]{Amir Levinson}
    \affiliation{The Raymond and Beverly Sackler School of Physics and Astronomy, Tel Aviv University, Tel Aviv 69978, Israel}

    \author[0000-0002-6987-1299]{Yuri Levin}
    \affil{Physics Department and Columbia Astrophysics Laboratory, Columbia University, 538 West 120th Street, New York, NY 10027}
    \affil{Center for Computational Astrophysics, Flatiron Institute, 162 5th Avenue, 6th floor, New York, NY 10010}
    \affil{Department of Physics and Astronomy, Monash University, Clayton, VIC 3800, Australia}

\begin{abstract}

We present the first numerical study of gravitational waves (GWs) from collapsar disks, using state-of-the-art 3D general relativistic magnetohydrodynamic simulations of collapsing stars. These simulations incorporate a fixed Kerr metric for the central black hole (BH) and employ simplified prescriptions for disk cooling. We find that cooled disks with an expected scale height ratio of $H/R\gtrsim0.1$ at $\sim10$ gravitational radii induce Rossby instability in compact, high-density rings. The trapped Rossby vortices generate vigorous coherent emission regardless of disk magnetization and BH spin. For BH mass of $\sim10\,M_\odot$, the GW spectrum peaks at $\sim100\,{\rm Hz}$ with some breadth due to various nonaxisymmetric modes. The spectrum shifts toward lower frequencies as the disk viscously spreads and the circularization radius of the infalling gas increases. Weaker-cooled disks with $H/R\gtrsim0.3$ form a low-density extended structure of spiral arms, resulting in a broader, lower-amplitude spectrum. Assuming an optimistic detection threshold with a matched-filter signal-to-noise ratio of 20 and a rate similar to Type Ib/c supernovae, LIGO--Virgo--KAGRA (LVK) could detect $\lesssim1$ event annually, suggesting that GW events may already be hidden in observed data. Third-generation GW detectors could detect dozens to hundreds of collapsar disks annually, depending on the cooling strength and the disk formation rate. The GW amplitudes from collapsar disks are $\gtrsim100$ times higher with a substantially greater event rate than those from core-collapse supernovae, making them potentially the most promising burst-type GW class for LVK and Cosmic Explorer. This highlights the importance of further exploration and modeling of disk-powered GWs, promising insights into collapsing star physics.
\end{abstract}

\section{Introduction}

The past decade has marked the beginning of the gravitational wave (GW) era, during which the LIGO-Virgo-KAGRA \citep[LVK;][]{Accadia2012,Acernese2015,LIGO2015,Abbott2018,Akutsu2021} collaboration has detected cosmic ripples arising from inspiraling sources \citep{Acernese2019,Tse2019,Buikema2020}. As we anticipate the upcoming phase with advanced LVK runs, followed by third-generation GW detectors such as Cosmic Explorer \citep[CE;][]{Abbott2017,Reitze2019,Evans2023}, Einstein Telescope \citep[ET;][]{Punturo2010a,Punturo2010b,Hild2011}, and Laser Interferometer Space Antenna \citep[LISA;][]{Babak2021}, there is a keen interest in discovering new GW sources, particularly those distinct from inspirals. The exploration of new types of GWs holds the promise of unveiling new insights into the physics of their progenitors and facilitating a deeper understanding of the diverse GW events occurring throughout the Universe.

Similar to inspiraling sources, burst-type sources are astrophysical events that generate transient GWs. However, their GW emission is less coherent, posing challenges for creating deterministic waveform models and, consequently, for their detection. A promising site for such burst-type sources is powerful explosions such as core-collapse supernovae (CCSNe). However, numerical simulations have shown that due to their quasi-spherical structure, only a minute fraction of the CCSN energy is released in GWs, where the dominant mechanism is stochastic excitation of the proto-neutron star (PNS) $f$- and $g$-modes, which are strongly coupled to gravitational radiation \citep[e.g.,][]{Nakamura2016,Andresen2017,Andresen2019,Andresen2021,Yakunin2017,Morozova2018,Radice2019,Abdikamalov2020,Mezzacappa2020,Mezzacappa2023,Vartanyan2023,Mezzacappa2024,Powell2024,Richardson2024}. The predicted small excitation amplitudes imply that even the third-generation GW detectors can only detect GWs powered by PNS accretion in the Milky Way and satellite galaxies at design sensitivity \citep{Srivastava2019,Evans2023,Corsi2024}.

If the progenitor star has sufficient angular momentum, its collapse will form an accretion disk around a newly formed black hole (BH; ``collapsar''). Accretion disks have been considered promising GW emitters due to various instabilities \citep{Kobayashi2003,vanPutten2003,Levinson2015,vanPutten2019}. If a global instability develops, nonaxisymmetric modes could give rise to density waves, introducing a quadrupole moment that emits GWs. One criterion to examine the stability of the disk is by comparing its pressure and gravity. The pressure gradient can counteract the gravitational forces trying to pull material inward, thereby preventing excessive collapse or disruption. If the disk becomes gravitationally unstable \citep{Chen2007}, it may fragment into high-density clumps \citep{Gammie2001}, which might give rise to strong GW emission \citep{Piro2007,Siegel2022}. However, it remains to be seen whether such accretion disks exist in astrophysical systems and the extent of their asymmetry, highlighting the need for general-relativistic magnetohydrodynamic (GRMHD) simulations.


Previous numerical attempts to assess the GW emission from accretion disks employed a preset self-gravitating torus setup \citep{Kiuchi2011,Korobkin2011,Wessel2021}. These studies found that nonaxisymmetric modes emerge in hydrodynamic simulations owing to Papaloizou-Pringle instabilities, which are quenched by magnetorotational instabilities (MRI) when magnetic fields are present \citep{Bugli2018,Wessel2023}. However, the relevance of torus studies to astrophysical systems such as collapsars is questionable for several reasons: (1) Ad-hoc prescribed tori evolve differently than accretion disks that form self-consistently during the collapse; (2) Torus simulations lack the constant gas supply onto the disk and the disk-envelope interaction; (3) Previous torus studies have not considered cooling effects, which are expected in the high-density collapsar disks owing to neutrino emission \citep[e.g.,][]{Narayan2001,vanPutten2003,Kohri2005,Chen2007,Just2022}. Cooling alters the disk thermodynamics and increases its density, potentially making it the primary driving mechanism for the disk to become gravitationally unstable \citep{Batta2014}, leading to vigorous GW emission.

Another potentially important source of density perturbations is the Rossby wave instabilities \citep[RWIs][]{Lovelace1999,Tagger1999,Li2000,Li2001}. RWIs emerge as a result of sharp gradients in the disk, which act as a potential well that can trap and amplify coherent structures like Rossby vortices. Therefore, RWIs are expected to be particularly strong in transiently forming disks. This insight motivates detailed first-principles simulations of self-consistently forming disks without considering self-gravity. Indeed, we find that the overdense Rossby vortices appear robustly in our simulations, creating a substantial mass quadrupole, responsible for generating GWs that may be detectable in LVK. The vigorous RWI-powered GWs were overlooked in previous studies because their setups lacked cooling mechanisms, necessary to increase the disk mass density and facilitate the formation of RWIs.

It is worth mentioning that there may be other mechanisms by which collapsars may produce strong GWs. For example, if a favorable magnetic field configuration is present, the central BH will power a relativistic jet that will generate a long gamma-ray burst \citep[lGRB;][]{LE93,Woosley1993}, and a hot turbulent cocoon, powered by the jet-star interaction. The aspherical nature and enormous energy of these outflows render the jet \citep{Sago2004,Birnholtz2013,Leiderschneider2021} and the cocoon \citep{Gottlieb2023a} interesting GW sources for LISA \citep{LISA2017} and CE/ET \citep{Gottlieb2023a}, respectively. While lGRB jets and cocoons are more promising GW emitters compared to quasi-spherical CCSNe, their scarcity reduces their GW detection prospects. Therefore, the detectable GW signals from accretion disks may be appreciably more abundant than those from relativistic outflows.

In this \emph{letter}, the first in a series of papers, we begin to build the case for serious and intense searches for GW bursts from collapsar disks by current and future GW interferometer teams. The structure of the letter is as follows. We outline the numerical setup and the GW calculation methodology in \S\ref{sec:numerical}. In \S\ref{sec:hydro}, we describe the disk evolution and the emergence of instabilities that give rise to nonaxisymmetric modes. Subsequently, in \S\ref{sec:gw_emission}, we present the resultant vigorous GW signal, followed by a discussion on the expected event rates in LVK and third-generation detectors in \S\ref{sec:detectability}. We deliberate on the implications of our results and conclude in \S\ref{sec:summary}.

\section{Numerical simulations}\label{sec:numerical}

\subsection{Numerical setup}

\begin{table*}[]
    \setlength{\tabcolsep}{3.0pt}
    \centering
    \renewcommand{\arraystretch}{1.2}
    \begin{tabular}{| c | c c c c c c c c | c c c c c | }
        
             \hline
        Model & Setup & $ H/R $ & $ \beta_p $ & $ {\rm max}(\sigma_0) $ & $ \abh $ & $ \mbh\,[\msun] $ & $ T_s\,[\rm s] $ & $ R_{\rm max}\,[r_g] $ & $ \egw\,[\erg] $ & $ \varrho $ (LVK) & $ \varrho $ (CE) & LVK rate [${\rm yr^{-1}} $] & $ \fgw\,[{\rm Hz}] $
        \\	\hline
        $ B $ & BNS merger & 0.1 & $ 10^4 $ & - & 0.68 & 2.67 & $ 0.3 $ & $ 10^3 $ & $ 2\times 10^{46} $ & 0.1; 0.2 & 0.4; 1.4 & 0 & 500-2000 \\   
        $ C $ & Collapsar & 0.1 & - & $ 10^{-3} $ & 0.8 & 10 & $ 24 $ & $ 10^5 $ & $ 7\times 10^{50} $ & 25; 46 & 390; 750 & $ \lesssim 1 $ & 30-300 \\ 
        $ Ca $ & Collapsar & 0.1 & - & $ 10^{-3} $ & 0.1 & 10 & 6.2 & $ 10^5 $ & $ 1.5\times 10^{49} $ & 16; 33  & 180; 360 & $ \lesssim 1 $ & 200-300 \\ 
        $ Cb $ & Collapsar & 0.1 & - & $ 10^{-4} $ & 0.8 & 10 & 7.4 & $ 10^5 $ & $ 8\times 10^{49} $ & 27; 54 & 350; 690 & $ \lesssim 1 $ & 100-200 \\ 
        $ Cc $ & Collapsar & 0.3 & - & $ 0 $ & 0.8 & 10 & 5.4 & $ 10^5 $ & $ 1.5\times 10^{49} $ & 5; 9  & 74; 130 & $ \lesssim 10^{-2} $ & 100-200 \\ 
            \hline
    \end{tabular}
    
    \caption{
        A summary of the models' parameters. The model names stand for the system type: BNS merger ($B$) and collapsar ($C$): $ C $ for our canonical setup; $ Ca $ for low $ \abh $; $ Cb $ for lower $ b $; and $ Cc $ for weaker cooling. $ H/R $ is the imposed disk scale height ratio which is effectively higher in models $ Ca $ and $ Cb $, $ \beta_p $ is the characteristic plasma beta in the disk, $ {\rm max}(\sigma_0) $ is the maximum initial magnetization in the star, $ \abh $ is the BH dimensionless spin, $ \mbh $ is the BH mass, $ T_s $ is the simulation time, and $ R_{\rm max} $ is the outer radius of the simulation grid in units of gravitational radius, $ r_g $. Quantities measured from simulations: $ \egw $ is the GW energy, and $ \varrho $ is the matched-filter SNR for edge-on and face-on observers, respectively, shown for LVK A+ and CE at $ D = 10\,{\rm Mpc} $, and corrected for accretion disk lifetime of $ t_{\rm acc} \approx 100\,\s $ (see \S\ref{sec:detectability}). The LVK rate is calculated assuming the model represents all collapsar disks and that their emergence rate is similar to SNe Ib/c rate. $ \fgw $ is the characteristic GW frequency range.
        }
        \label{tab:models}
\end{table*}

A robust GW signal may emerge from compact, dense accretion disks. Such conditions are prevalent in accretion disks around BHs formed owing to the gravitational collapse of a rotating stellar core. Similar conditions also exist in disks formed following a merger of two compact objects, at least one of which is a neutron star (NS). Both of these systems exhibit high mass accretion rates of $ \dot{M} \gtrsim 10^{-2}\,\msun $ s$^{-1}$, making them susceptible to neutrino cooling. Analytic models \citep{Chen2007} and numerical simulations incorporating neutrino transport \citep{Siegel2019,Foucart2023} suggest that neutrino-cooled disks cool down to a characteristic ratio of disk height to radius of $ 0.1 \lesssim H/Rx \lesssim 0.3 $, depending on the effective viscosity and BH spin. $ H/R $ tends to be smaller, $ H/R\approx 0.1 $, in the innermost disk regions, where, as we shall see, most of the GW radiation is emitted. Neutrino transport schemes are computationally expensive and thus do not allow following the disk evolution over long timescales. Therefore, our simulations do not incorporate full neutrino transport, but a cooling prescription that artificially maintains a fixed $ H/R $ \citep{Noble2009}. This prescription mimics radiative cooling by reducing the internal energy of the disk over the local Keplerian timescale until it matches a target disk scale height.

The fiducial collapsar simulations (model $C$ in table \ref{tab:models}) adhere to the setup outlined in \citet{Gottlieb2022b}, employing a stellar mass of $ M_\star = 20\,\msun $, and maximum initial magnetization ($ \sigma_0 \equiv b^2/4\pi\rho c^2 $, where $ b $ is the comoving magnetic field strength and $ \rho $ is the comoving mass density) within the stellar core capped at $ {\rm max}(\sigma_0) = 10^{-3} $. The angular momentum profile provides a fixed circularization radius of $ r_{\rm circ} = 25\,r_g $, similar to what is suggested by stellar evolution models during the first several seconds \citep{Gottlieb2024}. The disk scale height is imposed to maintain $ H/R = 0.1 $ at all times. The BH mass and dimensionless spin parameter are $ \mbh = 10\,\msun $, and $ \abh = 0.8 $, respectively. In addition to our fiducial model $ C $, we also consider a slowly-spinning BH with $ \abh = 0.1 $ (model $ Ca $) and a weaker $ B $-field (model $ Cb $). Both exhibit effectively thicker disks which increase in height from $ H/R = 0.1 $ close to the BH to $ H/R \gtrsim 0.2 $ at $ R \gtrsim 20\,r_g $, consistent with neutrino-transport collapsar simulations (Issa et al. 2024, in prep.). These disks are thicker because less gravitational energy (due to increased innermost stable circular orbit (ISCO) in $ Ca $) or magnetic dissipation ($ Cb $) is available to heat, making the cooling process less efficient. We also consider a weakly-cooled disk with $ H/R = 0.3 $ in model $ Cc $ which increases to $ H/R \approx 0.4 $ at $ r = 20\,r_g$, similar to a collapsar disk without cooling in \citet{Gottlieb2022a}. We find that GWs from weakly-cooled magnetized disks with $ H/R = 0.3 $ are outshone by the cocoon-powered GWs. Therefore, to investigate the GW signature of the weakly-cooled disk, we impose $ b = 0 $ in model $ Cc $ such that no electromagnetically-driven jet and cocoon form. We outline the list of simulations in Table~\ref{tab:models}.

In Appendix \ref{sec:mergers}, we present a post-merger simulation, which mirrors that conducted by \citet{Gottlieb2023e}. In short, it begins with a merger of NSs of masses $ 1.06\,\msun $ and $ 1.78\,\msun $, described by the LS220 equation of state for NSs \citep{Lattimer1991}. The merger product is a BH with a mass of $ \mbh = 2.67\,\msun $ and $ \abh = 0.68 $. Surrounding the BH is a substantial accretion disk weighing $ M_d = 0.096\,\msun $. The vector potential is defined by: 
\begin{equation}\label{eq:Aphi}
A = A_\phi \propto
\begin{cases}
\rho & \rho/\rho_{\rm max} \geq 0.2 \\
&\\
0 & \rho/\rho_{\rm max} < 0.2 \\
\end{cases} ~,
\end{equation}
and is imposed to be divergence-free. The potential normalization is chosen so that the gas-to-magnetic pressure ratio is $ \beta_p \approx 10^4 $.

We conduct the simulations using the 3D GPU-accelerated code \textsc{h-amr} \citep{Liska2022}, employing an ideal equation of state with an adiabatic index of $ 4/3 $, suitable for radiation-dominated gas. The simulations adopt spherical polar coordinates, $r$, $\theta$, $\phi$ with a logarithmic cell distribution along the $r$-direction, spanning from just inside the event horizon to $ R_{\rm max} $, which for collapsars is located outside of the star radius (see Tab.~\ref{tab:models}). Along the $\theta$- and $\phi$-directions, the cells are distributed uniformly. In total, the base grid comprises $N_r\times N_\theta \times N_\phi $ cells in the $r$-, $\theta$- and $\phi$-directions, respectively. For the collapsar and binary NS (BNS) merger simulations, $ N_r = 320, 192 $; $ N_\theta = 96, 96 $; and $ N_\phi = 192, 96 $, respectively. We incorporate local adaptive time-stepping and static mesh refinement. One (two) level of refinement is applied in the innermost $ 100\,(50)\,r_g$ in all collapsar (merger) simulations. We ensure that the wavelength of the fastest-growing MRI mode in the magnetohydrodynamic simulations is adequately resolved when the disk reaches a steady state evolution. In particular, we find that the MRI quality factor $Q_{\rm MRI} $, denoting the number of cells per the MRI wavelength, satisfies $ Q_{\rm MRI} \gtrsim 30 $ in both the $ \hat{\theta} $ and $ \hat{\phi} $ directions, surpassing the minimum requirement of $ Q_{\rm MRI} \sim 10 $ \citep[e.g.,][]{Hawley2011}. 

\subsection{GW approximations}

The GW fields can be computed using the Green's function method:
    \begin{equation}\label{eq:T_eff}
    h^{\mu\nu}(r,t) \approx \frac{4G}{c^4}\int{\frac{T_{eff}^{\mu\nu}\left(\pmb{r}',t-\lvert \pmb{r}-\pmb{r}'\rvert/c\right)}{\lvert \pmb{r}-\pmb{r}'\rvert}d^3r'}\,,
\end{equation}
where $T_{eff}^{\mu\nu} = \sqrt{-g}T^{\mu\nu} + t^{\mu\nu}$, $T^{\mu\nu}$ is the contravariant stress-energy tensor in Einstein equations, and $t^{\mu\nu}$ is a pseudo-tensor composed of nonlinear combinations of $ h_{\mu\nu} $ and their derivatives (see, e.g., Sec. V.A in \citealt{Thorne1980} for details). Eq.~\eqref{eq:T_eff} constitutes an integral equation for $h^{\mu\nu}$ and is therefore not practically useful for numerical computations. In the linear approximation, where internal stresses of the source can be neglected, Eq.~\eqref{eq:T_eff} can be rendered applicable by omitting the term $t^{\mu\nu}$.  However, this approach is invalid for self-gravitating systems, such as the accretion disks under consideration. For sufficiently slow systems, for which the reduced wavelength largely exceeds the GW source's characteristic size, a multipole emission expansion can be sought. In the weak gravity limit, wherein $t^{0\mu} \ll T^{0\mu}$, useful formulae for the mass quadrupole, mass octupole, and current quadrupole moments, that account for gravitational stresses, can be derived \citep[see][]{Fuller2013}. The mass quadrupole formula they obtained can be reduced, in this regime, to the standard form (see also 
\citealt{Misner1973} for a similar derivation) that we use
to compute the low-frequency emission [see Eq.~\eqref{eq:Qdot} below]. For the high-frequency emission, the multipole expansion fails because the reduced wavelength $\lambda/(2\pi)$ becomes comparable to the source size. We therefore have to use Eq.~\eqref{eq:T_eff} directly to compute the time-dependent perturbations to the metric. 
The right-hand side of this equation contains the pseudo-tensor representing gravitational stresses in the system; their direct computation goes beyond the scope of this paper. Instead, we use Eq.~\eqref{eq:T_eff} in the linear approximation and neglect $t^{\mu\nu}$.  We argue that this might be justified for unbound material, however, how accurate is the high-frequency spectrum thereby obtained remains to be assessed. It is reassuring however that the Green's function and the quadrupole approximation give consistent results in the intermediate range of frequencies, as we show in Appendix \ref{sec:comparison}.

As stated above, the multipole expansion approximation is only applicable when the reduced wavelength is larger than the size of the source along the line of sight, e.g., for edge-on observers:
\begin{equation}\label{eq:quad_crit}
2\pi\fgw R_d \lesssim c\,,    
\end{equation}
where $ R_d $ is the density wave orbital radius. For the angular frequency $ \Omega $ of the emitting gas in the quasi-Keplerian disk:
\begin{equation}\label{eq:omega}
    \Omega = \sqrt{\frac{G\mbh}{R_d^3}}\,,
\end{equation}
the characteristic GW peak frequency is:
\begin{equation}\label{eq:fGW}
    \fgw = \frac{\tilde{m}\Omega}{2\pi}\approx 180\tilde{m}\left(\frac{\mbh}{10\,\msun}\right)^{1/2}\left(\frac{100\,\km}{R_d}\right)^{3/2}\,{\rm Hz}\,.
\end{equation}
where $ \tilde{m} $ is the nonaxisymmetric mode. Eq.~\eqref{eq:fGW} implies that the multipole expansion remains valid when $ R_d \gtrsim \tilde{m}^2 r_g $. For the quadrupole moment ($ \tilde{m} = 2 $), this corresponds to $ \fgw \lesssim 800\,{\rm Hz} $.

\subsubsection{Comparison to CCSN studies}

Previous CCSN studies \citep[e.g.,][]{Nakamura2016,Andresen2017,Andresen2019,Andresen2021,Yakunin2017,Radice2019,Mezzacappa2020,Mezzacappa2023,Vartanyan2023,Powell2024,Richardson2024} have employed advanced numerical simulations to track the SN shock expansion. By applying the quadrupole approximation in post-processing, they have estimated the resulting GW signal. These studies have identified distinct GW signals at both low and high frequencies.

The high-frequency GW emission  ($ \fgw \gtrsim {\rm kHz} $) originates from accretion onto the PNS and convection within the PNS. The compactness of the PNS ensures that the quadrupole approximation remains valid even at high frequencies [Eq.~\eqref{eq:quad_crit}]. This was recently confirmed by \citet{Micchi2023}, who compared the high-frequency emission obtained by the quadrupole approximation with the Neuman-Penrose formalism and found that the two methods are consistent around the kilohertz peak. 

The low-frequency emission ($ \fgw \lesssim 0.2\,{\rm kHz} $) arises from the neutrino heating region below the SN shock wave. In these simulations, the shock wave expands to distances of $ \sim 10^3\,\km $, for which Eq.~\eqref{eq:quad_crit} dictates that the quadrupole approximation is only valid at $ f \lesssim 50\,{\rm kHz} $. The invalidity of the quadrupole approximation for the low-frequency peak calculation underscores the need for more reliable calculations to verify the low-frequency emission.


\subsubsection{Post-process calculation}

We solve the GWs from the simulations by combining two methods: The quadrupole approximation at frequencies below $ \sim 0.2\,{\rm kHz} $, 
and the flat spacetime Green's function at high frequencies (see Appendix \ref{sec:comparison}). For the quadrupole calculation, we follow \citet{Vartanyan2023} to calculate the first time-derivative of the quadrupole\footnote{To validate the quadrupole approximation, we verify that the octupole component in \citet{Fuller2013} can be neglected at $ f \lesssim 800\,{\rm Hz} $.}:
\begin{equation}\label{eq:Qdot}
    \dot{Q}_{m,ij}(r,t) = \int{\rho^{\rm ret}(v_i^{\rm ret}x_j+v_j^{\rm ret}x_i-\frac{2}{3}\delta_{ij}v_r^{\rm ret}r)dx_i'dx_j'dx_k'}\,,
\end{equation}
where all retarded quantities are defined as $ q^{\rm ret} = q\left(r',t-\lvert r-r'\rvert/c\right) $. Then, the strain tensor components are:
\begin{equation}\label{eq:hmunu}
    h_{\mu\nu}(r,t) \approx \frac{2G}{c^4D}\frac{d\dot{Q}_{m,\mu\nu}(r,t)}{dt}\,,
\end{equation}
where $ D $ is the distance to the source.

For the Green's function calculation, we follow \citet{Gottlieb2023a} who calculated the emission from unbound ejecta, neglecting gravity. The strain tensor components are determined using the far field limit of Eq.~\eqref{eq:T_eff} in flat spacetime, specifically:  
\begin{equation}\label{eq:h2}
    h_{\mu\nu}(r,t) \approx \frac{4G}{c^4D}\int{T_{\mu\nu}\left(\pmb{r'},t-\frac{\lvert \pmb{r}-\pmb{r'}\rvert}{c}\right)dx_i'dx_j'dx_k'}\,,
\end{equation}
where $ T_{\mu\nu} $ is the covariant stress-energy tensor, and the integration spans the volume of the disk.

Without the loss of generality, for an observer along the $ \hat{x} $-axis, the two independent solutions constituting the plane wave polarizations in the transverse-traceless (TT) gauge metric are given by:
\begin{equation}\label{eq:polarizations}
\begin{split}
    h_+ = h_{yy}-h_{zz}\,;\\
    h_\times = 2h_{yz}\,.
\end{split}
\end{equation}
Analogous solutions can be derived for observers situated along the $ \hat{y} $- and $ \hat{z} $-axes.

The characteristic strain is defined as \citep{Moore2015}:
    \begin{equation}\label{eq:hc}
        h_c(f) = 2f\lvert \tilde{h}(f)\rvert\,,
    \end{equation}
where
\begin{equation}\label{eq:fourier}
    \tilde{h}(f) = \sqrt{\tilde{h}_+^2(f)+\tilde{h}_\times^2(f)}\,,
\end{equation}
with $ \tilde{h}_\times(f) $ and $ \tilde{h}_+(f) $ representing the strain polarizations in the frequency domain.
To evaluate the signal detectability and event rate, we estimate the matched-filter signal-to-noise ratio (SNR) as \citep{Flanagan1998,Moore2015}:
\begin{equation}\label{eq:SNR}
    \varrho = \sqrt{\int_0^\infty \frac{4 |\tilde{h}_A(f)|^2}{S(f)}df}\,,
\end{equation}
where $ S(f) $ denotes the noise power spectral density of the GW detector, for which we use the \textsc{PyCBC} package~ \citep{Usman2016,Nitz2021} for the noise power-spectral densities. Parameter $ \tilde{h}_A(f) $ is the dimensionless GW strain, defined by:
\begin{equation}\label{eq:superposition}
        \tilde{h}_A(f) = F_+\tilde{h}_+(f) + F_\times \tilde{h}_\times(f)\,,
    \end{equation}
where $ F_+, F_\times $ are functions of the antenna pattern of the detector, which depend on the sensitivity to each polarization and the sky location of the observer. For the SNR calculations, we adopt $ F_+ = 1 $ and $ F_\times = 0 $.

\section{Hydrodynamic evolution}\label{sec:hydro}

    \begin{figure}
        \centering
        \href{https://oregottlieb.com/videos/Collapsar_cooled_disk.mp4}
        {\includegraphics[width=3.4in,trim={0cm 0cm 0cm 0cm}]{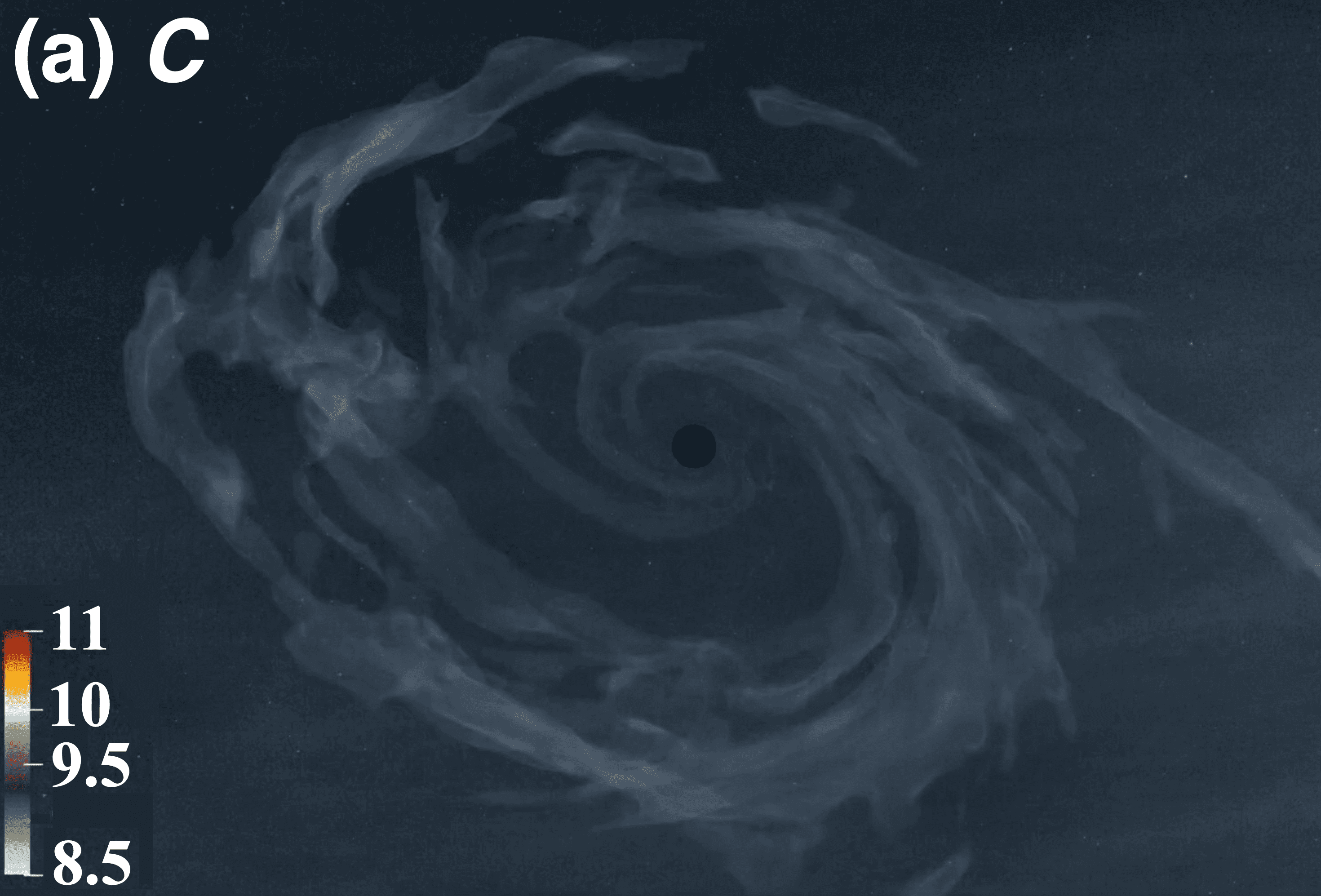}}
        \href{https://oregottlieb.com/videos/Collapsar_cooled_disk.mp4}
        {\includegraphics[width=3.4in,trim={0cm 0cm 0cm 0cm}]{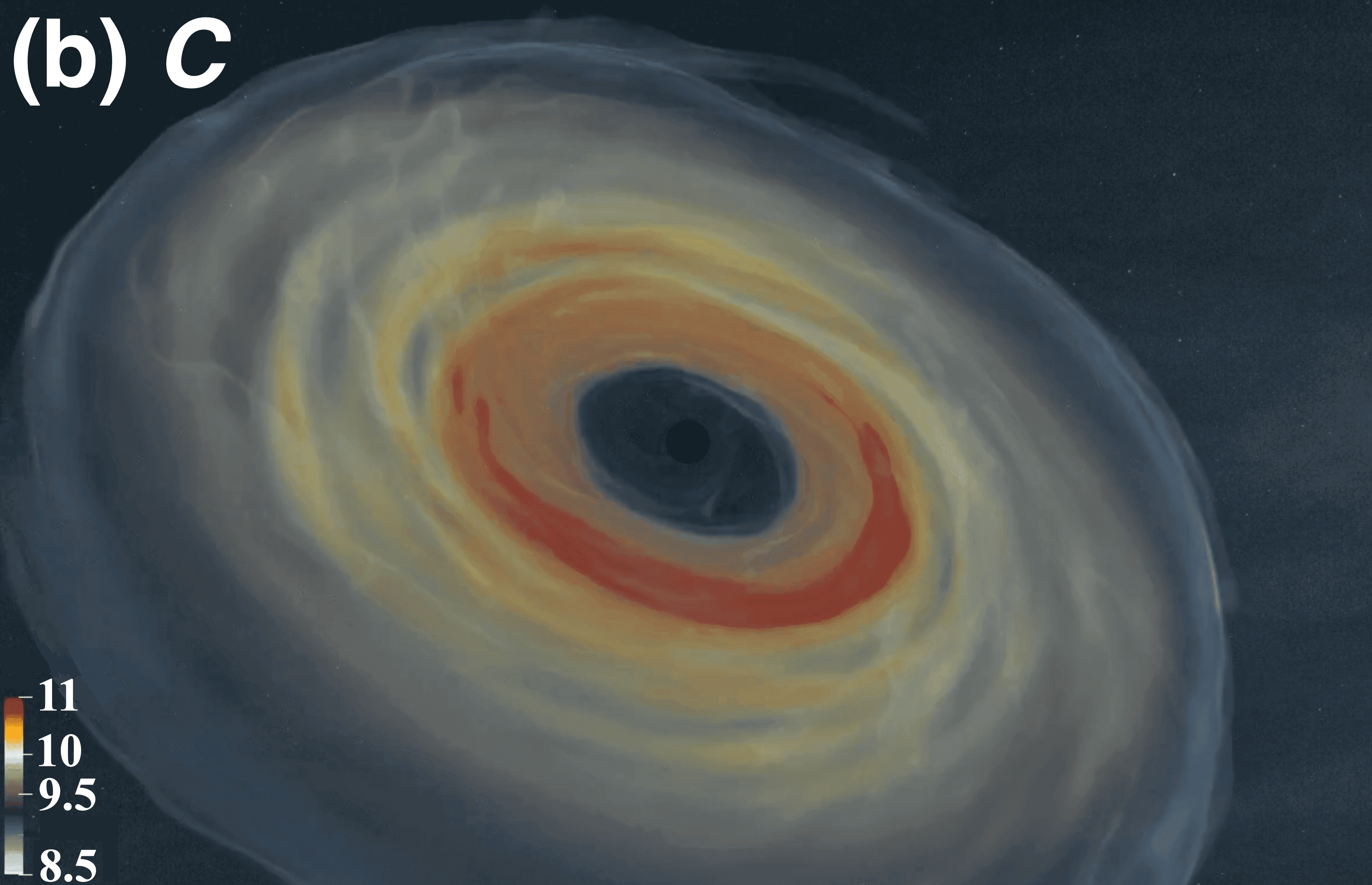}}
        \href{https://oregottlieb.com/videos/Collapsar_noncooled_disk.mp4}
        {\includegraphics[width=3.4in,trim={0cm 0cm 0cm 0cm}]{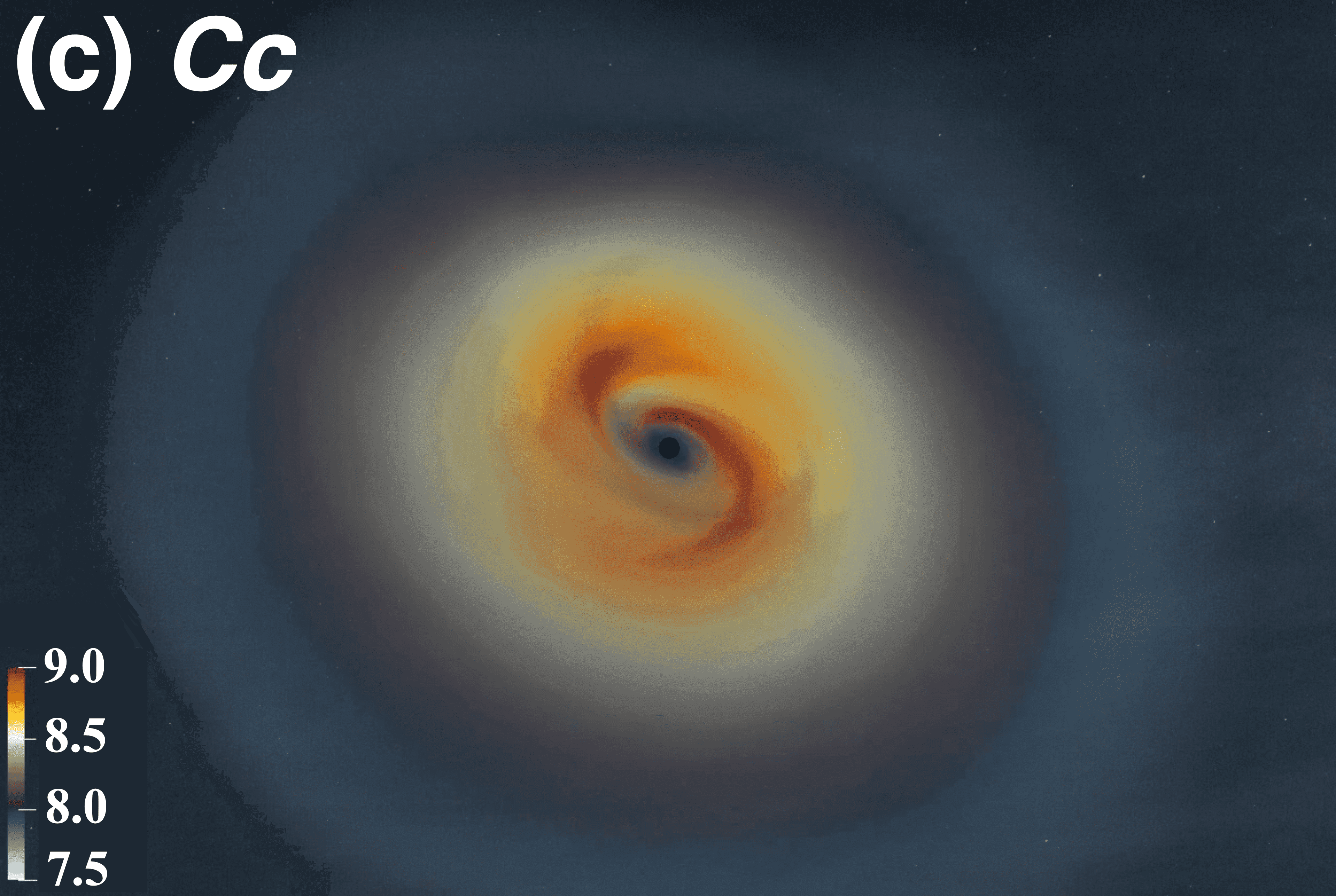}}
        \caption{3D renderings illustrate the logarithmic mass density (in .c.g.s.) of collapsar accretion disks. The length scale can be measured with respect to the BH event horizon radius, represented by a black circle. {\bf (a)} The early formation stages of the disk ($ \approx 1\,\s $) feature low-density spiral density waves. {\bf (b)} At $ t \gtrsim 1\,\s $, with the accumulation of infalling gas, a high-density ring forms at $ R \approx 15\,r_g $. RWIs in the disk give rise to nonaxisymmetric modes (shown at $ t \approx 5\,\s $). Over time, the disk viscously spreads and its density progressively drops. {\bf (c)} In the absence of strong cooling, the disk exhibits an extended structure [note that the horizon size in panels (a) and (b) is considerably larger than that in panel (c)], and attains lower densities compared to strongly-cooled disks. Consequently, weakly-cooled disks exhibit spiral arms at all times shown at $ t \approx 2\,\s $), giving rise to a weaker GW signal. Full movies showcasing the evolution of all collapsar accretion disks are available at \url{http://www.oregottlieb.com/disk_gw.html}.}
        \label{fig:3dc}
    \end{figure}

The first stellar shell with a circularization radius exceeding the ISCO radius, will form an accretion disk. We simulate the collapse of a spherically symmetric mass distribution star, incorporating a solid body rotation profile that facilitates the rapid formation of an accretion disk. Figure~\ref{fig:3dc} presents 3D renderings depicting the logarithmic mass density at different phases of strongly-cooled disk formation (panels a,b) and for weakly-cooled disks (c). The strongly-cooled accretion disks (models $ C, Ca, Cb $) exhibit similar evolution. We identify two stages in the formation of strongly-cooled disks in collapsars. As illustrated in Figure~\ref{fig:3dc}(a), during the initial stages of disk formation ($ \lesssim 1\,\s $), the disk is composed of low-density plasma. In low-density environments, pressure and density waves propagate freely, enabling differential rotation to amplify small perturbations into large, coherent spiral arms. Consequently, at the onset of disk formation, the radial expansion of the spiral waves results in a roughly flat and attenuated GW spectrum.

Fig.~\ref{fig:3dc}(b) indicates that over time, the accumulation of gas in the disk increases the disk density, leading to sharp pressure, density, and magnetic field gradients that render the disk Rossby unstable \citep[e.g.,][]{Tagger1999} -- The centrifugal force acts against the gradient to trap a pronounced high-density Rossby vortex within the disk. Thus, even without self-gravity in our simulations, the RWIs generate high-density ``banana''-shaped vortices that contain a substantial fraction of the disk's mass, as shown in Fig.~\ref{fig:3dc}(b). The stability of the flow in the disk against the development of turbulence due to shear and density gradients can be assessed through the Richardson number:
\begin{equation}
    R_i = g\left(\frac{1}{\gamma}\frac{d{\rm ln}p}{dr}-\frac{d{\rm ln}\rho}{dr}\right)\left(r\frac{d\Omega}{dr}\right)^{-2}\,,
\end{equation}
where $ \gamma $ is the adiabatic index, and $ g $ denotes the effective gravity:
\begin{equation}
    g = \frac{GM}{r^2}-r\Omega^2\,.
\end{equation}
If $ R_i \lesssim 0.25 $, the flow becomes unstable. We find that in our simulations $ R_i < 0 $ in the high-density regions in the innermost $ \sim 20 r_g $, supporting the conditions for the development of local turbulence. The turbulent mixing and differential rotation may facilitate the formation of sharp gradients, which are essential for the emergence of Rossby vortices. We leave a detailed analysis of the RWI criterion in collapsar disks for future work.

    \begin{figure*}
        \centering
        \includegraphics[width=3.2in,trim={0cm 0cm 0cm 0cm}]{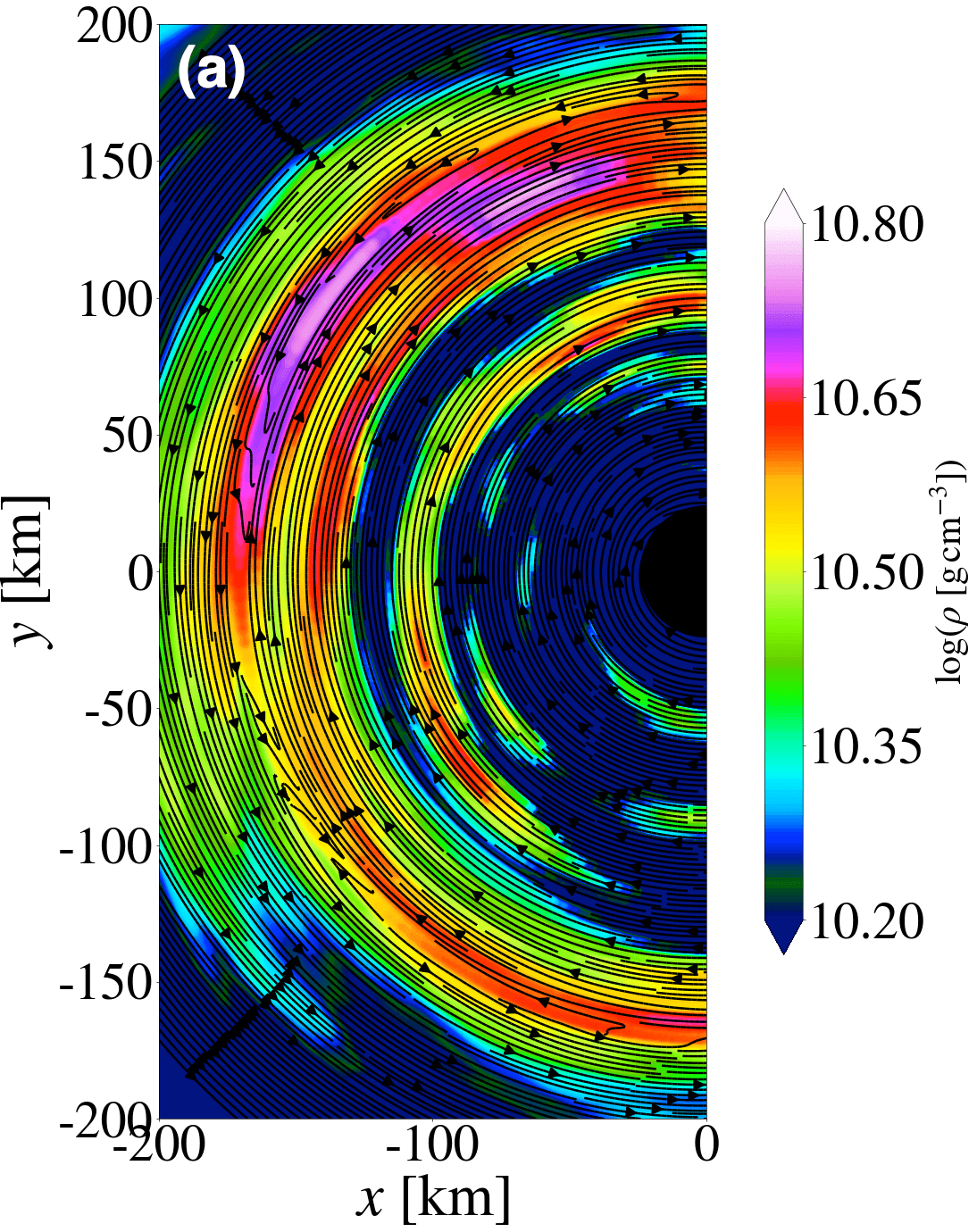}
        \includegraphics[width=3.2in,trim={0cm 0cm 0cm 0cm}]{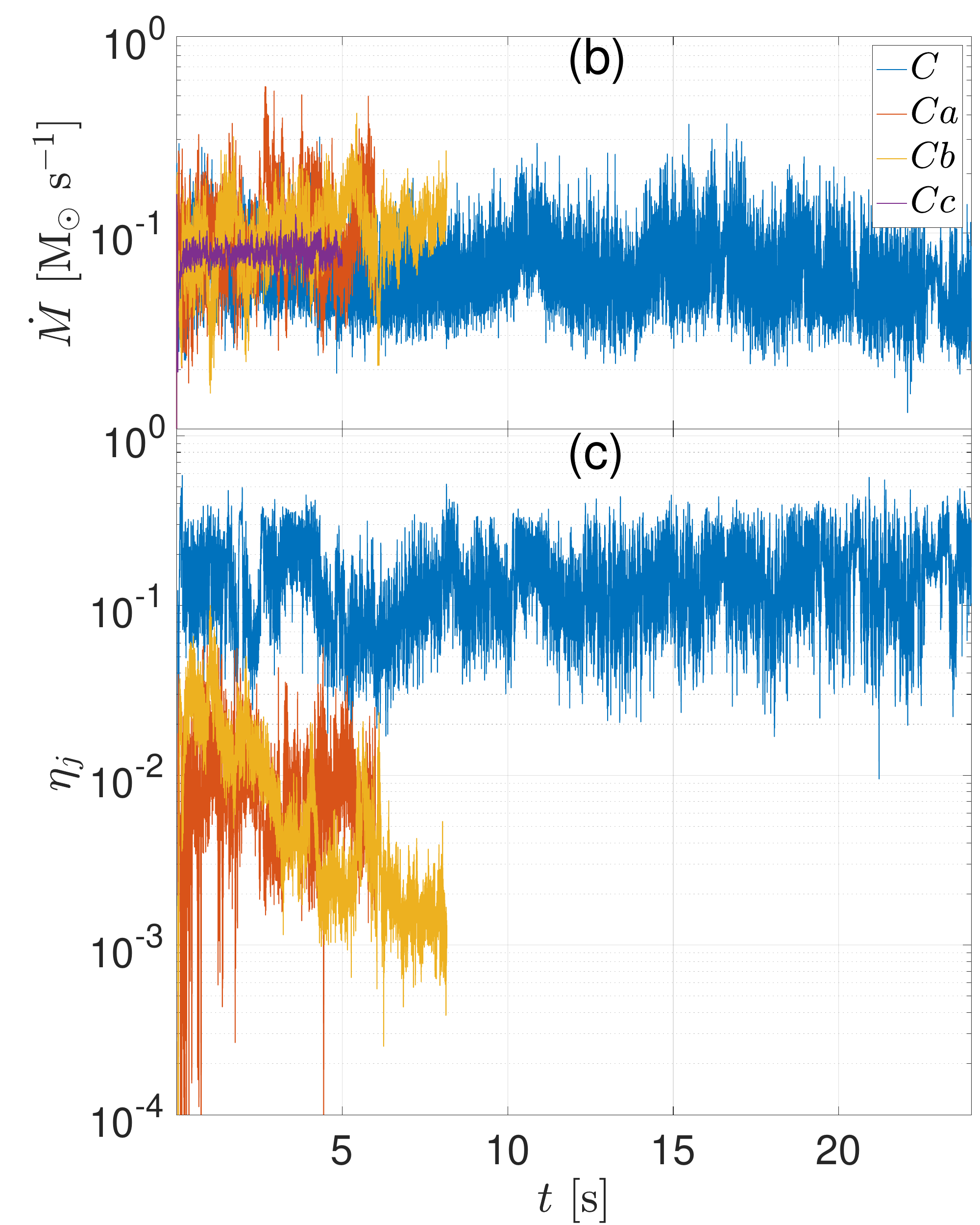}
        \caption{{\bf (a)} Logarithmic mass density map of model $ C $ at $ t=3.4\,\s $ with velocity streamlines in the corotating frame of the high-density wave at $ (x,y) \approx (-150,100)\,\km $ portray a trapped Rossby vortex. {\bf (b)} The mass accretion rate on the BH horizon remains roughly constant in time and across models. {\bf (c)} The jet efficiency on the BH horizon is substantial for rapidly spinning BHs with strong magnetic fields.}
        \label{fig:accretion}
    \end{figure*}

Figure~\ref{fig:accretion}(a), which presents the velocity streamlines in the disk, offers visual evidence of these high-density clumps behaving as Rossby vortices within their circular orbit. Namely, in the corotating frame, the vortex flow consists of banana-shaped closed streamlines and it is trapped between prograde and retrograde circular flows on either side of the vortex. These modes emerge stochastically, with the $ \tilde{m} $-mode number constantly changing throughout the simulation (see movies at \url{http://www.oregottlieb.com/disk_gw.html}).

Stellar evolution models suggest that the circularization radius remains roughly constant during the first several seconds at $ R \sim 20\,r_g $ \citep{Gottlieb2024}, as employed in our simulations. Nevertheless, strong magnetic fields facilitate MRI, which transports angular momentum outwards, thereby driving disk expansion and reducing disk density. This implies that later falling gas will merge with the outer parts of the disk before reaching its circularization radius. After several seconds, stellar evolution models \citep[e.g.,][]{Woosley2006} suggest that the circulation radius of the gas increases as a function of the radius at the time of collapse, and thus the disk expansion will become faster. We leave a detailed exploration of the impact of the stellar rotational profile on the disk structure for future work.

Fig.~\ref{fig:3dc}(c) depicts the extended accretion disk in model $ Cc $, characterized by weak cooling with $ H/R = 0.3 $. The increased disk size results in lower mass density in the disk, similar to the hydrodynamic conditions at early times of strongly-cooled disks. The low-density conditions in the disk set the stage for developing the spiral arm structure with mode $ \tilde{m} = 2 $.

Fig.~\ref{fig:accretion}(b) indicates that all models exhibit a comparable constant mass accretion rate on the BH, as expected from free-falling gas with an initial radial power-law index of $ -1.5 $ \citep{Gottlieb2023b}. The subtle differences between the models stem from the outflow feedback on accretion, with stronger outflows impeding accretion more. Fig.~\ref{fig:accretion}(c) depicts the jet efficiency on the BH horizon, $ \eta_j $, defined as:
\begin{equation}
    \eta_j = \dot{M}^{-1}\int_{r_h, \sigma>1}\sqrt{-g}(-T^r_t-\rho u^r)\mathrm{d}\theta \mathrm{d}\varphi\,,
\end{equation}
where the integration is calculated on the BH horizon $ r_h $, and only for fluid elements with magnetization $ \sigma > 1 $, $ g $ is the metric determinant, $ T^r_t $ denotes the radial energy flux density component of the mixed stress-energy tensor $ T $, and $ u^\mu $ is the four-velocity such that $ \rho u^r $ represents the radial mass-energy flux density. Model $ C $ exhibits high jet efficiency through the Blandford-Znajek mechanism \citep[BZ;][]{Blandford1977}. In contrast, slowly rotating and weakly-magnetized BHs can eject merely $ \eta_j \lesssim 1\% $ of the accreted power owing to a smaller angular momentum reservoir or weaker magnetic flux on the horizon.   We note that the emergence of collapsar jets in weakly-magnetized disks is observed only when cooling is strong \citep[see][for the absence of jets in weakly-magnetized thick disks]{Gottlieb2022a}.
\section{GW emission}\label{sec:gw_emission}

\subsection{Analytic estimates}

We begin by providing an analytic approximation for the characteristics of the GW signal. Given that $ 2\pi\fgw R_d/c=m\sqrt{r_g/R_d} \lesssim 1 $ at the peak frequency, we can employ the gravitational quadrupole moment ($ \tilde{m} = 2 $) to estimate the GW strain magnitude as:
\begin{equation}\label{eq:h}
    h = \frac{2G}{Dc^4}\frac{d^2Q_m}{dt^2} \approx \frac{2G}{Dc^4}{E_d\epsilon}\,,
\end{equation}
where $ \epsilon $ signifies the degree of asymmetry introduced by the azimuthal modes and $ E_d $ is the disk energy, estimated as:
\begin{equation}\label{eq:Ed}
    E_d \approx \frac{1}{2}\frac{G\mbh M_d}{R_d} \approx 10^{52}\frac{\mbh}{10\,\msun}\frac{M_d}{0.1\,\msun}\frac{100\,\km}{R_d}\,\erg\,,
\end{equation}
where $ M_d $ is the mass of the disk. We find the disk mass and radius in model $ C $ to grow linearly with time such that $ E_d \approx {\rm const} \approx 7\times 10^{51}\,\erg $. Plugging Eq.~\eqref{eq:Ed} into Eq.~\eqref{eq:h}, the characteristic strain of the disk can be estimated as:
\begin{equation}\label{eq:hd}
    h_c \approx 7\times 10^{-23}\epsilon\frac{10\,{\rm Mpc}}{D}\frac{\mbh}{10\,\msun}\frac{M_d}{0.1\,\msun}\frac{100\,\km}{R_d}\,.
\end{equation}
Since $ \mbh \sim R_d $, $ h_c $ is independent of the BH mass and scales as $ M_d $, which depends on the disk thickness, the circularization radius of the gas and the envelope mass.

Comparing Eq.~\eqref{eq:hd} with our numerical calculations, we find that $ \epsilon \approx 0.3 $ in model $ C $ and consistently $ \epsilon \gtrsim 0.2 $ in all strongly-cooled disks, but $ \epsilon \approx 0.1 $ when cooling is weak in model $ Cc $. Suppose the high-density wave dominating the GW emission is in a quasi-Keplerian orbit. In that case, the characteristic GW peak frequency is governed by the angular frequency $ \Omega $ [Eq.~\eqref{eq:omega}] and the $ \tilde{m} $-mode, given by Eq.~\eqref{eq:fGW}. Higher subdominant modes (multipoles) will broaden the spectrum above the peak frequency. As larger orbits emerge with the disk expansion, the emission will shift to lower frequencies. If the high-density wave exhibits radial dependence, such as in spiral waves, it will span varying distances from the BH. This implies that the emission originates from different orbits, resulting in a broader spectrum and a weaker GW signal \citep[see also][]{Levinson2015,vanPutten2019}.

\subsection{Simulation results}

    \begin{figure*}
    \centering
        \href{https://oregottlieb.com/videos/GW_polarizations.mp4}
        {\includegraphics[width=7in,trim={0cm 0cm 0cm 0cm}]{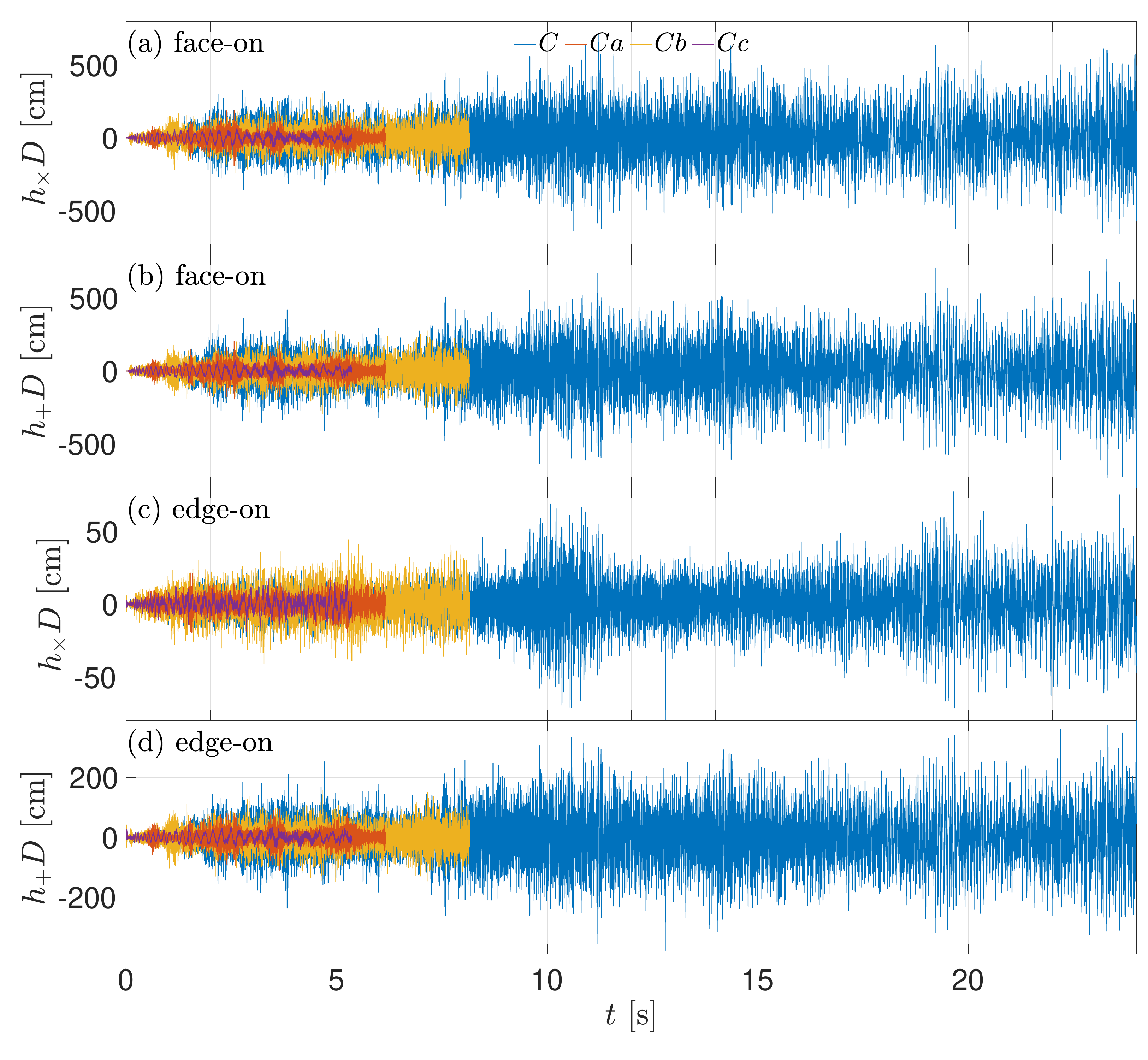}}\\
        \caption{
        Cross {\bf (a,c)} and plus {\bf (b,d)} polarizations for face-on observers {\bf (a,b)} and edge-on observers {\bf (c,d)} feature two characteristic timescales. Short timescales, characterized by high variability, are dictated by the orbit of nonaxisymmetric density waves close to the BH. Modulations on longer timescales ($ \sim 0.1\,\s $) in model $ Cc $ imply that modes at larger orbits contribute to the GW spectrum in extended disks.
     }
     \label{fig:polarization}
    \end{figure*}

Figure~\ref{fig:polarization} depicts the two GW polarizations observed by face-on (a, b) and edge-on (c, d) observers, as described in Eq.~\eqref{eq:polarizations}. All polarizations feature high variability stemming from the orbital timescale of the high-density waves. As discussed earlier, the amplitude of the short timescale variability is lower in model $ Cc $, where $ H/R=0.3 $, compared to the strongly-cooled disks with $ H/R \gtrsim 0.1 $. Additionally, in model $ Cc $, all observers would detect a lower frequency peak over $ \sim 0.1\,\s $ seen by the sinusoidal waves (purple lines). This phenomenon arises from larger orbits in the extended weakly-cooled disk.

We observe a phase shift of $ \pi/4 $ between cross and plus polarizations for face-on observers (i.e., circular polarization), and an anti-phase between cross and plus polarizations for edge-on observers (linear polarization). As we shall see next, the GW spectra are rather similar for different observers. This suggests that detecting the elliptical polarization could serve as a probe for the inclination angle of the disk relative to Earth's line of sight. We plan to explore this further in future work.

    \begin{figure*}
    \centering
        \href{https://oregottlieb.com/videos/Spectrum.mp4}
        {\includegraphics[width=7.in,trim={0cm 0cm 0cm 0cm}]{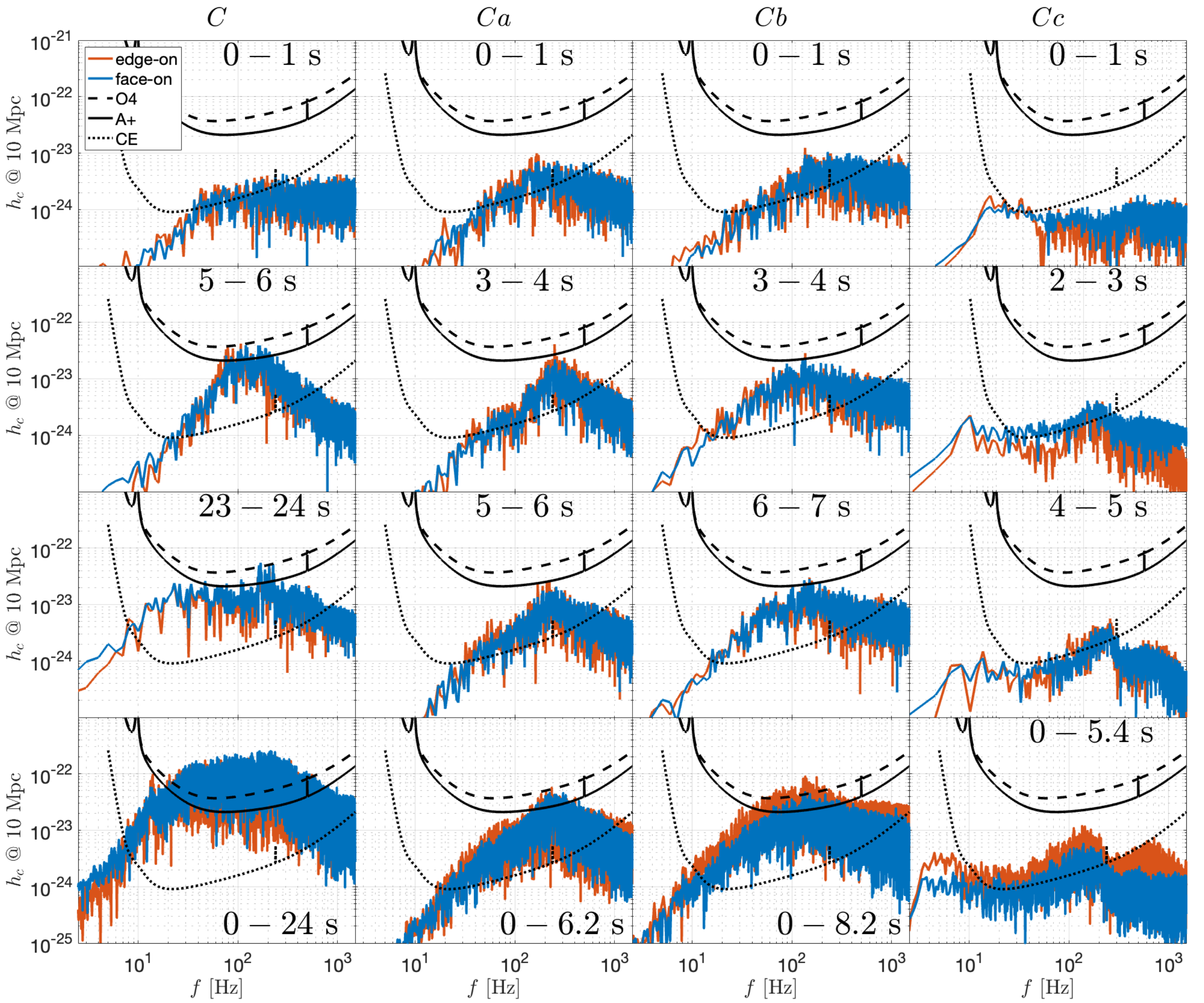}}
        \includegraphics[width=2.3in,trim={0cm 0cm 0cm 0cm}]{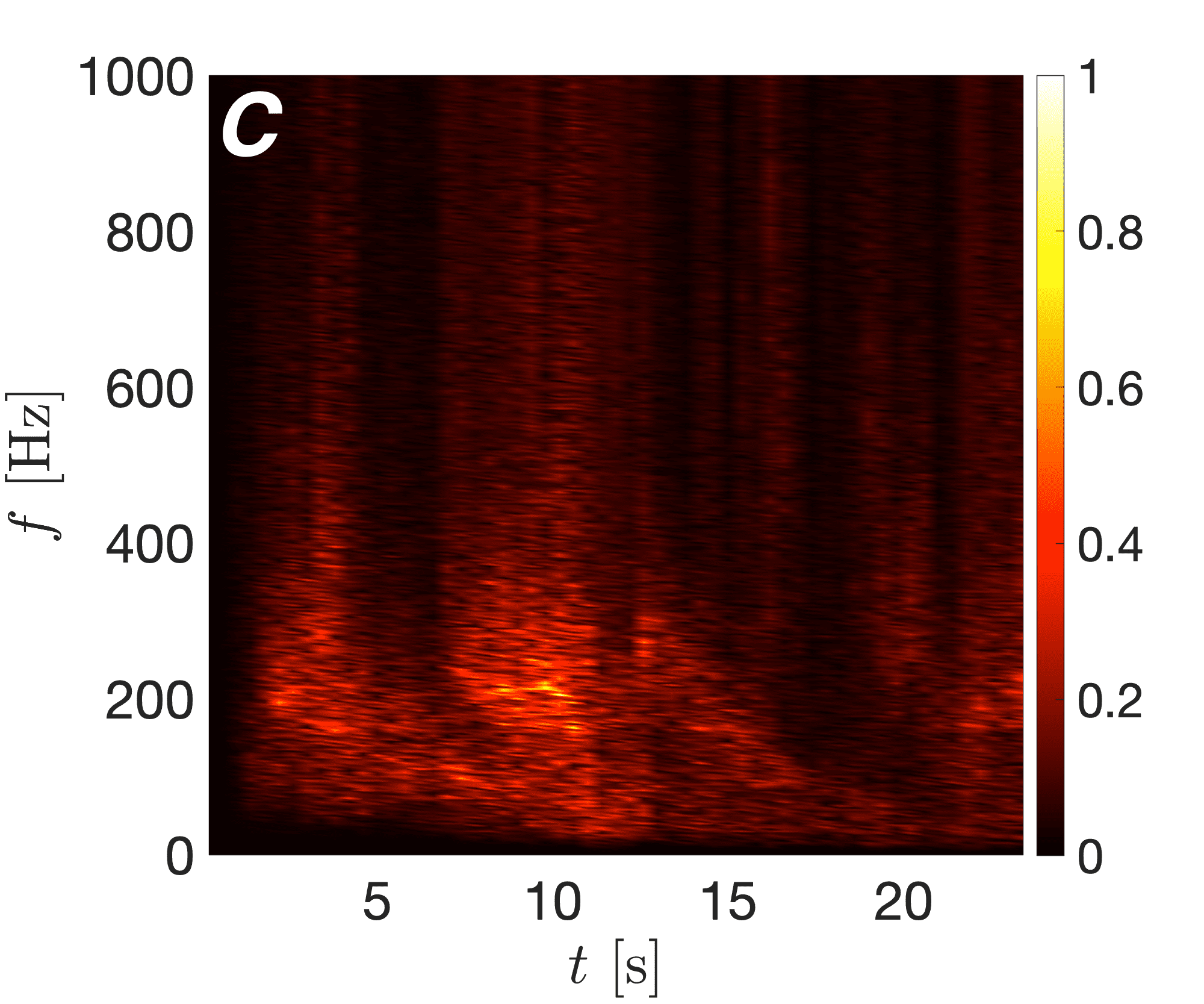}
        \includegraphics[width=2.3in,trim={0cm 0cm 0cm 0cm}]{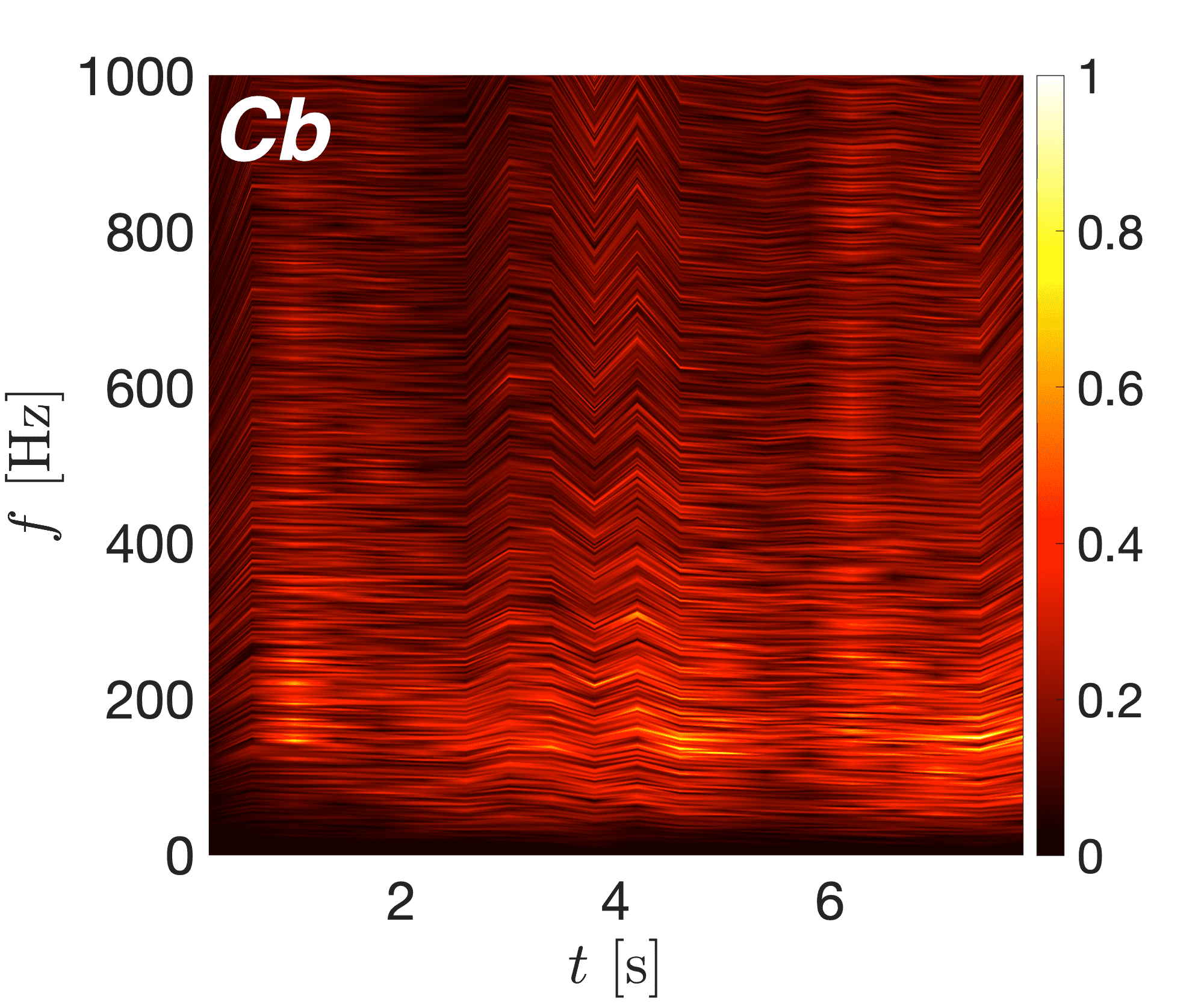}
        \includegraphics[width=2.3in,trim={0cm 0cm 0cm 0cm}]{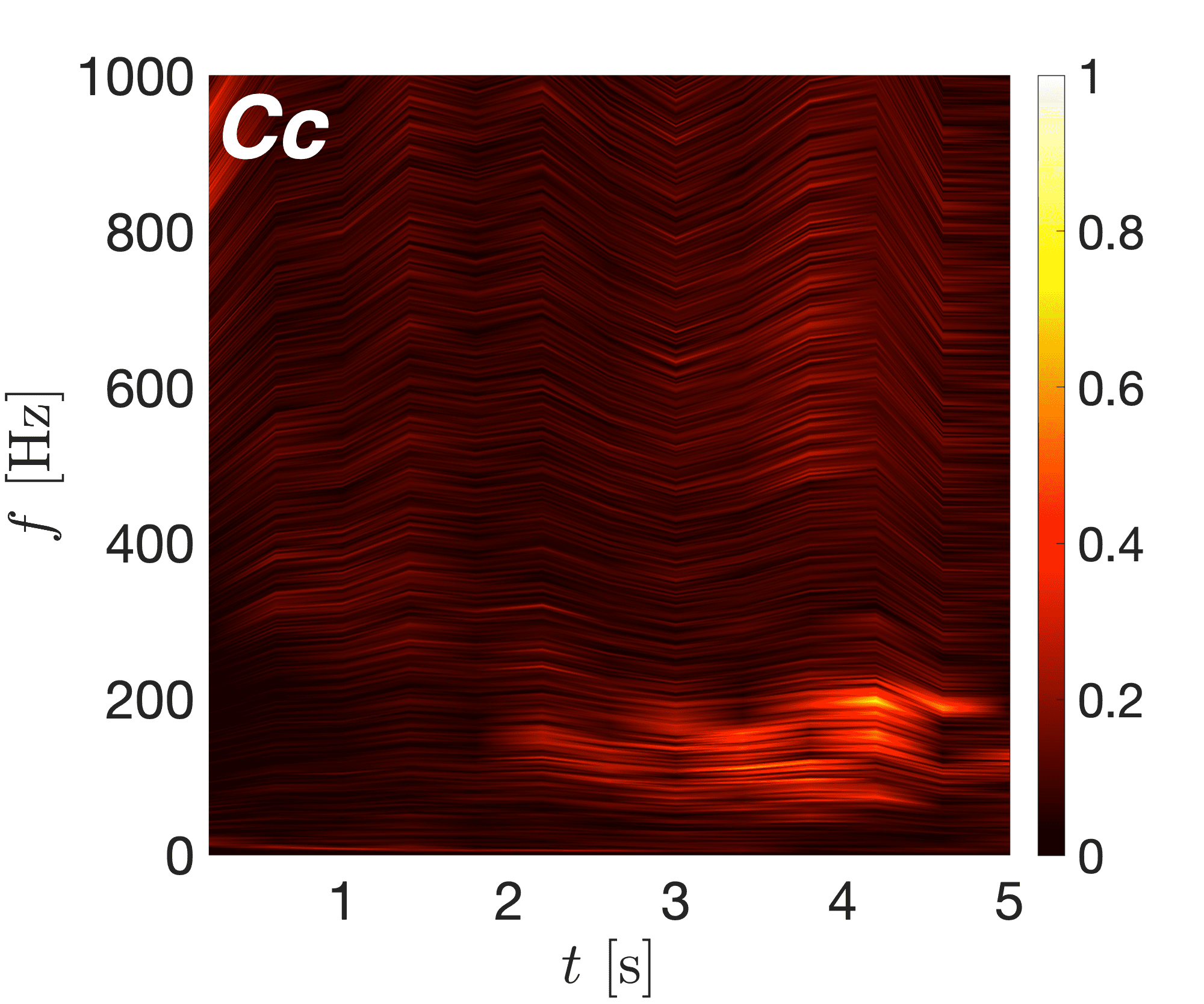}
        \caption{
        {\bf Top}: edge-on (red) and face-on (blue) spectra for models $ C $ (first column from left), $ Ca $ (second column), $ Cb $ (third column), and $ Cc $ (fourth column) are presented, with different rows depicting the spectra at various times. The black lines represent the sensitivity of various GW detectors. During the disk formation stage (first row), all models exhibit a quasi-flat spectrum. When cooling is strong (first three columns), Rossby vortices develop, leading to pronounced GW peak frequencies as the disk evolves. A movie showcasing the evolution of the GW spectrum is available at \url{http://www.oregottlieb.com/disk_gw.html}.
        {\bf Bottom}: The spectrograms for face-on observers over a moving time-window of $ 0.4\,\s $ illustrate the signal amplification after $ \lesssim 1\,\s $. The spectrogram of model $ Ca $ resembles that of $ C $.
     }
     \label{fig:collapsar}
    \end{figure*}

Figure~\ref{fig:collapsar} depicts the integrated GW spectra at various time windows for the different models. During the disk formation stage within the first $ \sim 1\,\s $ [Fig.~\ref{fig:3dc}(a)], low-density spiral waves predominantly contribute to the GW emission. These spirals, spanning various radii, translate to a roughly flat GW spectrum, characterized by a relatively weak signal (first row in Fig.~\ref{fig:collapsar}). Over time, accretion proceeds, the disk mass grows and the GW emission intensifies. For strongly-cooled disks, high-density rings form at $ R_d \approx 15\,r_g $ [Fig.~\ref{fig:3dc}(b)]. Instabilities within these rings give rise to azimuthal modes induced by high-density waves [Fig.~\ref{fig:accretion}(a)], which yield coherent GW emission. Eq.~\eqref{eq:fGW} suggests that during this phase, the GW signal peaks at $ \sim 100\,\tilde{m}\,{\rm Hz} $. As $ \tilde{m} $ changes stochastically over time, higher multipoles contribute to the spectra some breadth at $ f \gtrsim 100\,{\rm Hz} $.

Comparing the GW spectra of strongly cooled disks, we find that all such disks evolve similarly, regardless of their magnetic field strength and BH spin. For example, models $ C $ and $ Ca $ exhibit similar evolution during the first $ \sim 6\,\s $, as shown by their spectra at $ 5\,\s < t < 6\,\s $. Over time, the disk viscously expands and the spectrum broadens to lower frequencies, as seen in the spectrum of model $ C $ at $ 23\,\s < t <24\,\s $, resulting in a broader spectrum in model $ C $ owing to its longer $ T_s $.

\begin{figure}
    \centering
        {\includegraphics[width=3.6in]{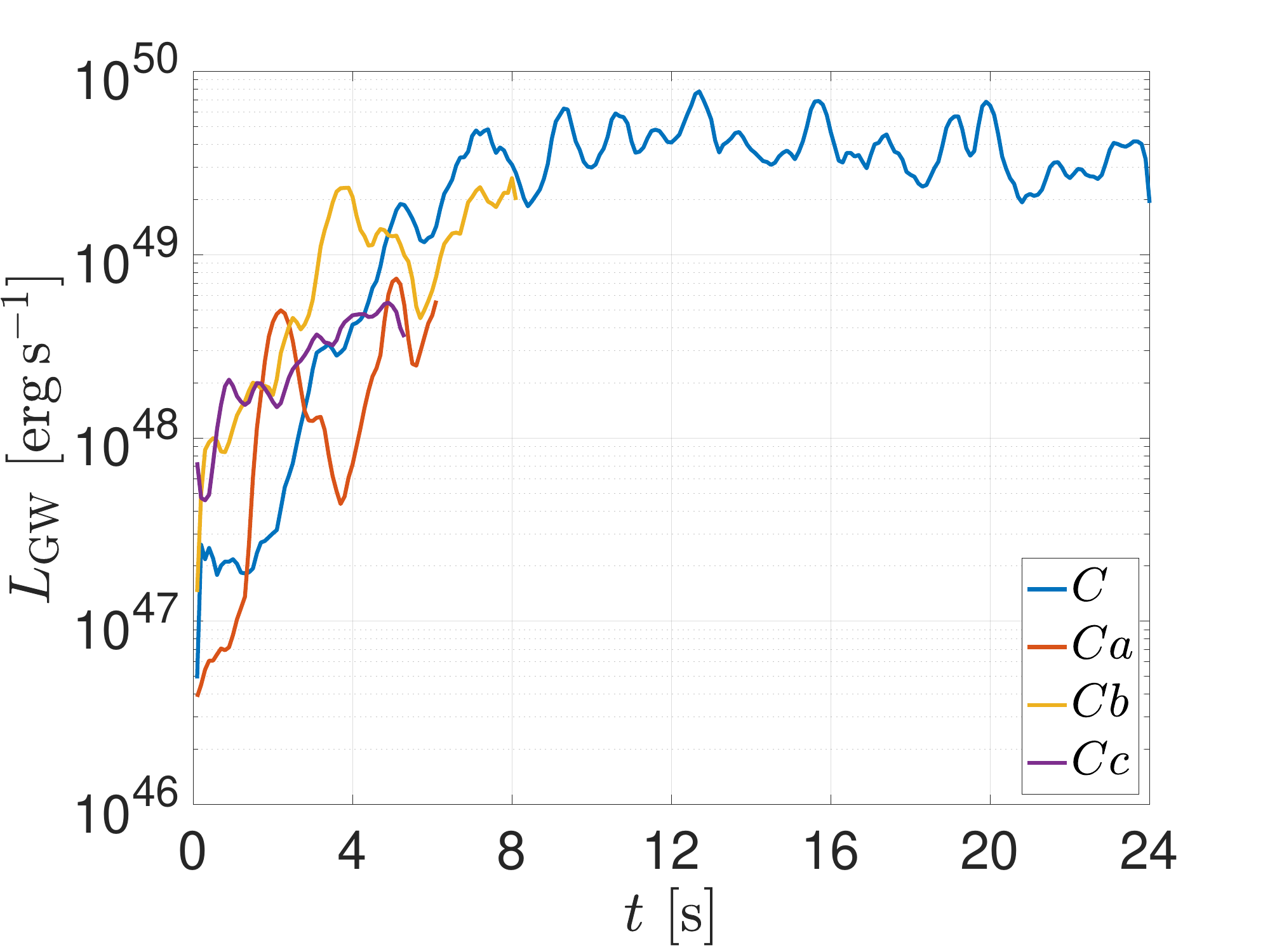}}
        \caption{The GW luminosity as a function of time, $ L_{\rm GW} = dE_{\rm GW}/dt $, illustrates that the GW power grows similarly in all models.}
     \label{fig:LGW}
    \end{figure}

We estimate the GW energy using:
\begin{equation}\label{eq:E_GW}
    \egw = \sum_{\mu\nu}\frac{c^3}{20G}\int_0^t\left(\frac{dh_{\mu\nu}D}{dt}\right)^2dt~.
\end{equation}
Fig.~\ref{fig:LGW} illustrates that all models exhibit a similar GW luminosity evolution in time, indicating that the differences in $ \egw $ in Tab.~\ref{tab:models} and in the integrated spectra over $ T_s $ between the models are attributed to the differences in $ T_s $. The increase in the GW luminosity can be explained by the accumulation of additional mass in the disk, which both increases the GW amplitude [Eq.~\eqref{eq:hd}] and facilitates the emergence of nonaxisymmetric modes. The asymptotic value of $ L_{\rm GW} \gtrsim 10^{49}\,\erg\,\s^{-1} $ suggests that a few percent of the total disk energy is radiated away as GWs. In nature, the lifetime of collapsar disks might be $ \sim 100\,\s $, on the order of lGRB timescales, implying the integrated GW energy may reach $ \egw \sim 10^{51}\,\erg $, as indicated by model $ C $ (see Tab.~\ref{tab:models}).

In model $ Cc $, characterized by weak cooling, some nonaxisymmetric modes emerge, similar to other models. However, this model does not feature a distinct high-density ring but spiral arms (\S\ref{sec:hydro}), resulting in a weaker integrated signal, also increasing linearly with time. In model $ Cc $, the disk extends to $ \sim 10^3\,\km $. Therefore, low-frequency modes are also present owing to the extended disk orbital frequency. The weakly-cooled disks have amplitudes consistent with what has been found by \citet{Wessel2023} for non-cooled disks.

The bottom panels of Fig.~\ref{fig:collapsar} depict the spectrograms in models $ C, Cb, Cc $. In strongly-cooled disks, the azimuthal density wave at $ R_d \approx 200\,\km $ emerges at $ t \gtrsim 2\,\s $, giving rise to peak GW frequencies at $ \fgw \approx 100-300\,{\rm Hz} $. The stochastic emergence of various nonaxisymmetric modes and the disk spreading result in a broad spectrum. In the weakly-cooled disk, no high-density rings are present, and the disk is more extended. This leads to a broader frequency range with lower characteristic frequencies, which are not fully captured in the spectrogram moving time windows of $ 0.4\,\s $.

With the strong GW emission in cooled disks, one may expect that the GW losses could alter the system dynamics. We make a crude estimate of this effect by treating the orbiting vortex as a member of a binary, with the central BH being the other member, and by computing the radiation-reaction timescale for such a binary. The latter is given by \citep[e.g.,][]{Shapiro1983}:
\begin{equation}\label{eq:radiation_timescale}
    t_{rr} = \frac{5}{256}\frac{c^5}{G^3}\frac{a^4}{M^2\mu}\,,
\end{equation}
where $ a $ is the semi-major axis, well-approximated by the orbital radius of the vortex, $ M \approx M_{\rm BH} $ is the total mass of the binary, is well-approximated by the BH mass, and $ \mu $ is the reduced mass of the binary, well-approximated by the vortex mass $m$. We find:
\begin{equation}\label{eq:rr_disk}
    t_{rr} = \frac{5}{256}\frac{c^5}{G^3}\frac{R^4}{\mbh^2m}\approx 10\left(\frac{10\,\msun}{\mbh}\right)^2\frac{10^{-2}\,\msun}{m}\left(\frac{R}{150\,\km}\right)^4\,\s\,.
\end{equation}
Gravitational radiation reaction is important if it acts on a shorter timescale than the viscous timescale, which is given by:
\begin{equation}\label{eq:tv}
    t_v = \frac{R^2}{\alpha c_s H}\approx 0.75\frac{R_d}{150\,\km}\left(\frac{\alpha}{0.1}\frac{c_s}{10^9\,{\rm cm\,\s^{-1}}}\frac{H/R}{0.2}\right)^{-1}\,\s\,,
\end{equation}
where $ \alpha $ is the effective viscosity coefficient and $ c_s $ is the sound speed.

The normalization values in Eqs.~\eqref{eq:rr_disk}, \eqref{eq:tv} are based on typical values found in the simulations. The order-of-magnitude ratio suggests that the gravitational radiation reaction does not dramatically alter the disk dynamics and the Rossby vortices. The $ \alpha $ viscosity parameter is not calculated in the simulation and its value can vary by more than an order of magnitude. However, when the magnetic fields are strong, as considered in collapsars, its value is typically $ \alpha \gtrsim 0.1 $ \citep{Kotko2012,Ju2017}, thereby unlikely to affect this result.

\section{Detectability}\label{sec:detectability}

Reliable computations of the SNR, a customary measure of detectability in GW astronomy, encounter significant uncertainties stemming from various factors:
(1) The accuracy of our findings is contingent upon our simplistic prescription of disk cooling;
(2) The absence of capturing global instabilities driven by self-gravity in our simulations presents a challenge to the robustness of the observed disk hydrodynamics;
(3) Although our simulations are the longest 3D GRMHD collapsar simulations ever performed, they are constrained to only a fraction of the expected disk's lifetime. Consequently, late-time disk characteristics may diverge from early-time conditions, e.g., due to non-linear feedback from outflows;
(4) The disk's lifetime and spatial dimensions are intricately linked to the progenitor's rotational and density profiles, which are poorly constrained observationally and fraught with uncertainties in theoretical modeling, particularly regarding angular momentum transport within rotating stars. Similarly, the masses of the stellar envelope and the BH affect $ M_d $ and $ R_d $, and thus the GW amplitude and frequency. Unlike the progenitor profiles, these quantities are inferred from observations but span a wide range of values.
(5) The fraction of stellar core collapses that lead to the formation of accretion disks remains poorly constrained observationally.

Advanced simulations should be designed to assess the uncertainty arising from factors 1 and 2. We incorporate the remaining uncertainties using the following approximations:

(3, 4) The lifespan of an accretion disk could be deduced from the rotational profile in stellar evolution models of rapidly rotating stars. Rapid rotation leads to the stripping of the outermost stellar envelope, resulting in a star radius of $ R \approx 10^{11}\,{\rm cm} $ \citep{Woosley2006}, corresponding to an accretion timescale of hundreds of seconds. However, the launch of outflows impedes accretion, shortening the accretion disk timescale. Our best estimate is that the simulated accretion disk lifetime is approximately an order of magnitude shorter than what simple infall time calculations suggest.

The expected disk lifetime is longer than our current simulation runs, and it would be appealing to introduce a simple scaling of SNR with $\tacc$. However, the increase in SNR with $ \tacc $ is not straightforward to infer. As shown in Fig.~\ref{fig:LGW}, the GW luminosity grows with mass accumulation in the disk over time. On the other hand, as the disk expands, the GW emission is anticipated to weaken and shift towards lower frequencies, where the sensitivity of LVK is diminished and that of CE increases [see Eq.~\eqref{eq:hd}]. The frequency is also contingent on $ \mbh $ and $ R_d $ as per Eq.~\eqref{eq:fGW}. A BH could potentially form with a mass as low as $ \mbh \approx 3\,\msun $, resulting in a smaller disk radius and higher frequencies -- $ \fgw \sim \mbh^{-1} $ [see Eq.~\eqref{eq:fGW}], deviating from the peak sensitivity range of both LVK and CE. Nevertheless, this might be offset by smaller BHs exhibiting an extended accretion timescale, which scales as $ \tacc \sim \mbh^{-0.5}$. Ultimately, the BH mass growth with $ \dot{M} \approx 0.1\,\msun\,\s^{-1} $ [Fig.~\ref{fig:accretion}(b)], and the increasing orbital radius due to the accretion of shells with larger circularization radius and the disk viscous spreading will augment $ R_d $, likely culminating in a signal peaking at $ \fgw \approx 100\,{\rm Hz} $, unless BH-driven outflows impede accretion in later stages. We consider a conservative potential enhancement of the numerically calculated SNR values [Eq.~\eqref{eq:SNR}] at $ D = 10\,{\rm Mpc} $ by a factor of $ 100\,\s/T_s $, given that $ \tacc $ might be hundreds of seconds while the SNR grows sub-linearly. Overall, collapsar disks power a GW signal with SNR in LVK (CE) of dozens (hundreds) at $ D = 10\,{\rm Mpc} $.

(5) We constrain the formation rates of collapsar accretion disks by establishing lower and upper limits. For the lower limit, we argue that as the formation of lGRB jets necessitates the presence of both an accretion disk and a robust magnetic field, collapsar disks are at least as common as lGRBs. The observed rate of the latter is $ {\cal{R}_{\rm lGRB}}\sim 1\,{\rm Gpc^{-3}}~{\rm yr}^{-1} $ \citep{Wanderman2010}, and thus for a typical jet opening angle of $ \theta_j \approx 0.1 $, the collapsar rate is $ {\cal{R}_{\rm Collapsar}}\approx 100 ~{\rm Gpc^{-3}}~{\rm yr}^{-1} $. It is worth noting that if the jet exhibits wobbling behavior, the inferred lower limit on the rate could be lower by up to an order of magnitude \citep{Gottlieb2023a}. For the upper limit, it has been proposed that most jets do not generate typical lGRBs \citep{Bromberg2012}, but rather power a separate class of low-luminosity GRBs (llGRBs) driven by a mildly-relativistic shock breakout \citep{Tan2001,Waxman2007,Nakar2015}. Such mildly-relativistic shocks necessitate the formation of jets, which are likely associated with collapsar disks. Therefore, we adopt the high llGRB rate $ {\cal{R}_{\rm llGRB}}\approx 10^4 ~{\rm Gpc^{-3}}~{\rm yr}^{-1} $ \citep{Soderberg2006}, which is also comparable to the rate of stripped-envelope Type Ib/c SNe (SNe Ib/c) of $ {\cal{R}_{\rm Ib/c}}\approx 2.6\times 10^4 ~{\rm Gpc^{-3}}~{\rm yr}^{-1} $, as an upper limit on the rate of collapsar disks. 

Following the above approximations and assuming that the simulated collapsar configurations represent the broader collapsing star population, we are ready to discuss the detection prospects of GWs from collapsar disks. Setting an optimistic matched-filter SNR threshold for detection at $ \varrho_c = 20\varrho_{20} $ (estimating $ 20 \lesssim \varrho_c \lesssim 50 $; Macquet, private communication), we project that LVK O4 holds the potential to detect GWs emanating from accretion disks up to distances of a few dozen Mpc. Conversely, CE is anticipated to discern such signals at distances extending to hundreds of Mpc. In the CE scenario, even weakly-cooled disks hold detectability prospects spanning dozens of Mpc.

If collapsar disk rates are similar to those of lGRBs, then the expected LVK event rate is $ \sim 10^{-2}\,\varrho_{20}^{-3}\,{\rm yr}^{-1} $. However, if disks emerge at rates similar to SNe Ib/Ic, a few GW events might already have been detectable by LVK, as indicated by more than a dozen SN Ib/c events within $ D \lesssim 40\,{\rm Mpc} $ in the Zwicky Transient Facility \citep[ZTF;][]{Graham2019,Masci2019}, including events as close as $ D \approx 15\,{\rm Mpc} $ \citep[e.g.,][]{Jacobson2020,Kilpatrick2021,Rho2021,Ho2023,Kuncarayakti2023}. In CE, several events per year might be detectable even if the cooling efficiency is low and collapsar disks are as common as lGRBs. If a substantial fraction of SNe Ib/c gives rise to accretion disks, hundreds of events could become detectable with third-generation GW detectors, depending on the disk cooling efficiency. In either scenario, both the strength and the rate of detectable GW signals from accretion disks are orders of magnitude more promising than those expected from CCSNe.

Electromagnetic counterparts have the potential to significantly enhance temporal and spatial localization, thereby aiding the search for GWs emanating from collapsar disks. Relativistic outflows such as lGRBs can provide valuable constraints on time localization, up to the propagation of the jet in the stellar envelope of $ \lesssim 100\,\s $. However, the detectability of such jets is often low, estimated at only $ \lesssim 1\% $ owing to their beamed emission \citep{Frail2001}. Non-relativistic electromagnetic counterparts are anticipated in most SN Ib/c events but offer weaker time localization, spanning from hour-long shock breakout to week-long radioactive decay emission post-collapse. Consequently, collapsars may not benefit from precise time localization, presenting a challenge in conducting targeted searches for GW emission from collapsar disks. Nevertheless, regardless of whether the electromagnetic signal aids in the detection of the GW signal, all GW events are expected to be accompanied by a detectable electromagnetic counterpart, at least those detectable by present-day GW detectors, rendering these events multi-messenger phenomena.

\section{Discussion}\label{sec:summary}

In this \emph{letter}, we underscore the potential of collapsar accretion disks as promising isotropic GW sources, even for present-day GW detectors. To investigate this, we conducted the first numerical simulations tracking the evolution of collapsar disks. We explored various disk magnetizations, BH spins, and cooling strengths to assess their respective impacts on the emitted GWs. We found a consistent picture across simulations with cooling exerting the most significant influence on the resulting GW emission. While post-merger disks are also subject to strong cooling, their relatively short lifetimes and low masses render their detection prospects unlikely (see Appendix \ref{sec:mergers}).

We implemented artificial cooling that operates instantaneously. Specifically, the cooling timescale is nearly zero, implying that cooling directly induces disk instability rather than through an increase in surface density \citep{Gammie2001}. When cooling is strong, the dominant instability observed is the Rossby-Wave Instability, which generates a Rossby vortex within a high-density ring in the innermost part of the disk at $ R_d \approx 15\,r_g $ \citep[see e.g.,][]{Tagger1999}. The number of high-density clumps dictates the azimuthal mode number, consequently governing the GW frequency. In our simulations, the characteristic GW frequency begins with $ \fgw \gtrsim 100\,{\rm Hz} $ with some breadth due to contributions from various azimuthal modes. As time progresses, the disk expands via viscous spreading and the larger circularization radius of the stellar outer shells, gradually shifting the spectrum towards lower frequencies. When the disk cooling is weaker, the large radial extent of the disk reduces the mass density in the disk. This supports the development of spiral arms, giving rise to a broader and attenuated GW spectrum. In \S\ref{sec:detectability}, we discussed potential uncertainties in the progenitor rotational profiles, but our understanding is that the qualitative behavior of the Rossby-Wave Instability should persist regardless of the specific model, provided that the cooling is sufficiently strong.

Our findings of a robust signal motivate the exploration of GWs from collapsar accretion disks in observed LVK data. \citet{Abbott2021} conducted an extensive all-sky search for long-duration ($ t \gtrsim 1\,\s $) GW transients in LVK O3 run data. They established upper limits on the GW strain for various transient sources, including accretion disks. Using our simulated spectra, we estimate these limits correspond to weak constraints on the non-detection of accretion disks at $ D \lesssim 1\,{\rm Mpc} $. However, previous searches did not utilize templates but focused on detecting excess power in spectrograms \citep{Thrane2011}. Consequently, accurately assessing the appropriate matched-filter SNR for detection remains challenging. Our conservative yet optimistic estimate of a matched-filter SNR of $ \varrho_c = 20 $ for detection corresponds to a maximum detection distance of a few dozen Mpc for GWs from collapsar disks.


In addition to uncertainties in $ \varrho_c $, the unknown rate with which collapsing stars form accretion disks presents a challenge in estimating the expected detection rate of these objects. Assuming that a significant fraction of stripped-envelope SNe is accompanied by disk formation and $ \varrho_c = 20 $, a rough estimate of the detection rate suggests that a few events could already be detectable using present-day GW detectors, highlighting the need for follow-up modeled and targeted searches. The number of GW detections from collapsar disks is anticipated to increase by orders of magnitude in future detectors such as Cosmic Explorer or Einstein Telescope. Even if the GW emission from collapsar disks is lower than our estimates, e.g. due to a lower cooling efficiency, dozens of events are expected to be detectable. To explicitly state our worst-case scenario expectations: if accretion disks form at a relatively low rate, closer to that observed for lGRBs, or a large fraction of them is not strongly concentrated and cooled, we may need to await future advancements in GW detectors to detect these GW events.

Collapsar accretion disks have emerged as highly promising sources of GW signals, boasting significantly greater strength by $ \gtrsim 2 $ orders of magnitude compared to extensively studied CCSN events. While exhibiting a comparable strength to cocoon-powered GWs, accretion disks are considerably more abundant, increasing their detection prospects. Furthermore, among all suggested GW sources from collapsars, accretion disks are anticipated to exhibit the most coherent pattern, potentially offering an opportunity to develop phenomenological templates for their GW bursts.

Detecting GWs emanating from accretion disks presents a novel opportunity to study the inner workings of collapsing stars, as the disk time evolution reflects the radial structure of the collapsing star. The duration of the GW signal holds valuable insights into the accretion timescale dictated by the balance between the collapse and the outflows. Furthermore, the GW spectrum can help determine important disk properties such as mass, radial extent, and energy. The GW elliptical polarization may reveal the orientation of the disk with respect to the line of sight. As statistical data accumulate, this approach aims to elucidate the intricate relationship between accretion disks, CCSNe, jet launching, and lGRBs.

In summary, considering the relatively high expected event rate, we argue that collapsar accretion disks might be the most promising burst-type GW sources to date. We therefore urge both the theoretical and observational communities to embark on exploration and modeling of GWs originating from collapsar accretion disks. On the theoretical side, enhancing the numerical experiments with neutrino-cooling schemes and self-gravity, as well as utilizing progenitors from stellar evolution models, will place theoretical predictions on firmer ground. On the observational side, it seems important to comprehensively characterize GW signals from various collapsar models and develop detailed templates for these signals. The GWs might be imminently detectable in current-day detectors and could be targeted by their electromagnetic counterparts. Such efforts will facilitate a more efficient GW search and enrich the depth of information attainable from a multi-messenger signal concerning the physical properties of the source.

\acknowledgements

We thank Will Farr, Anthony Mezzacappa, David Radice, Katerina Chatziioannou, Eric Thrane, Neil Cornish, Adrian Macquet, Florent Robinet, Adam Burrows, and Sharan Banagiri for useful discussions.
OG is supported by the Flatiron Research Fellowship.  This work was supported by a grant from the Simons Foundation (MPS-EECS-00001470-04) to AL and YL. YL's work on this subject is also supported by Simons Investigator grant 827103.

\bibliography{refs}

\appendix

\section{Compact binary mergers}\label{sec:mergers}

\begin{figure*}
    \centering
        \href{https://oregottlieb.com/videos/Post_merger_disk_GW.mp4}
        {\includegraphics[width=3.82in]{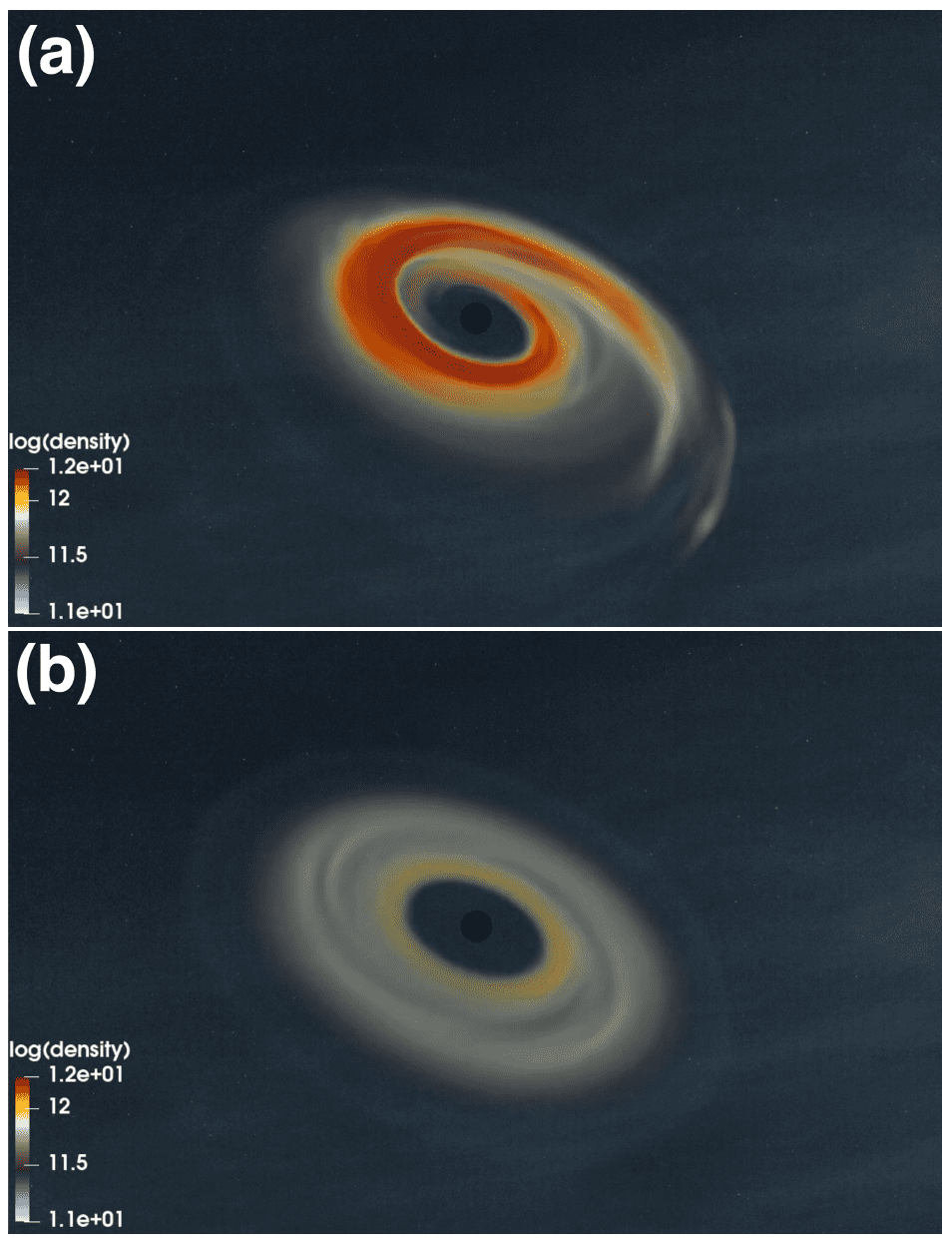}}
        \includegraphics[width=3.05in]{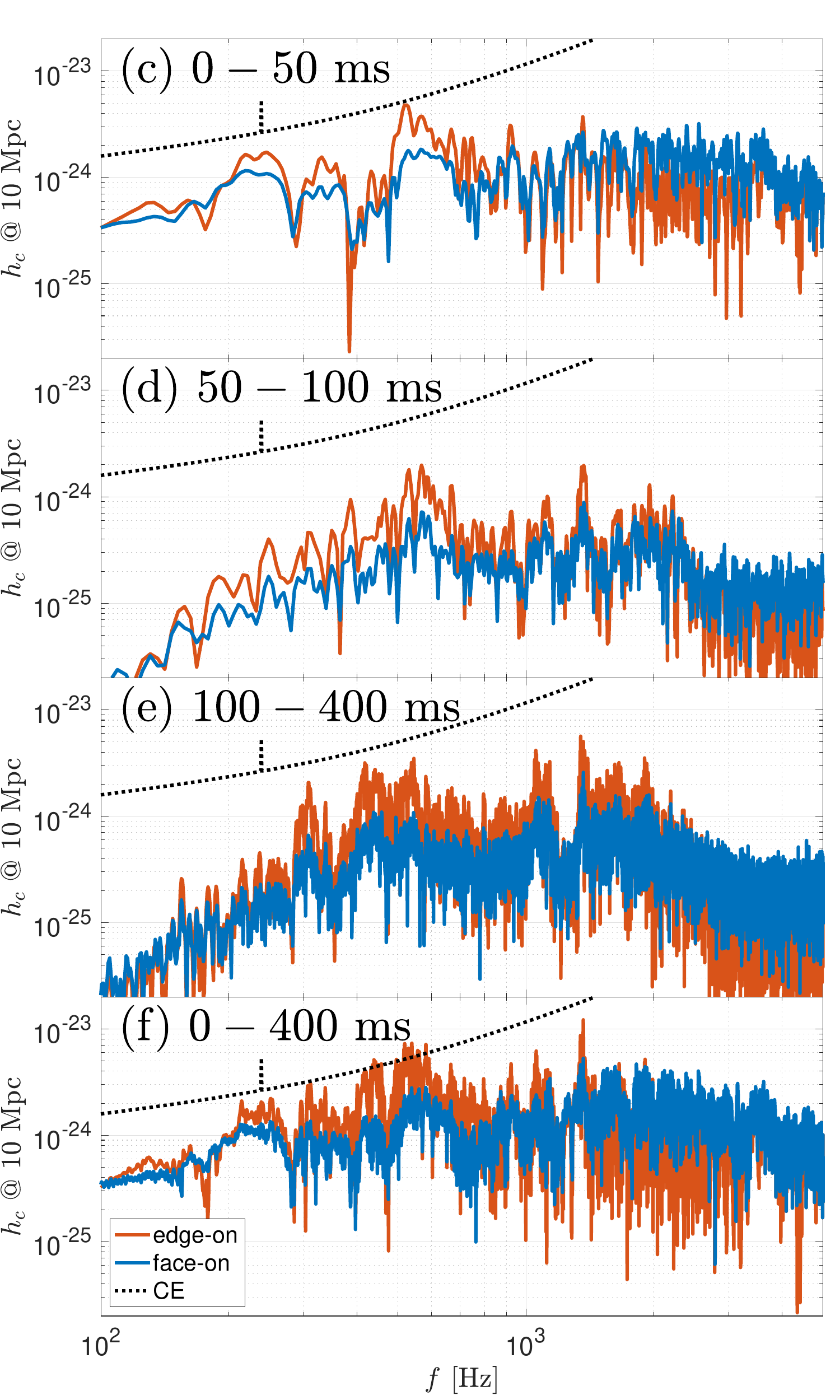}
        \caption{3D renderings illustrate the logarithmic mass density (in .c.g.s.) of post-merger accretion disks. The length scale can be measured with respect to the size of the BH in the center. {\bf (a)} The early formation stages of the disk ($ \sim 10\,{\rm ms} $) feature high-density spiral arms resulting from the disrupted NS. {\bf (b)} As the matter axisymmetrized, the compact accretion disk becomes homogenous. The full movie showcasing the evolution of the post-merger accretion disk is available at \url{http://www.oregottlieb.com/disk_gw.html}.
        {\bf (c-f)} edge-on (red) and face-on (blue) spectra for model $ B $, with different rows depicting the spectra at various times. The dotted black lines represent the CE sensitivity curve.}
     \label{fig:bns}
    \end{figure*}

During the merger of a BNS or a BH--NS system, some of the NS crust is disrupted and circularized around the merger remnant. The tidal disruption of the NS naturally gives rise to nonaxisymmetric modes. This phenomenon is illustrated in Figure~\ref{fig:bns}(a), which presents a 3D rendering of the logarithmic mass density around the post-merger BH $ 10\,{\rm ms} $ after a BNS merger. The spiral structure persists for a few dozen milliseconds before transitioning into an axisymmetric configuration within a more uniformly distributed post-merger disk [Fig.~\ref{fig:bns}(b)]. The compact nature of these post-merger disks \citep{Fernandez2020} indicates that their characteristic GW frequency is notably higher compared to those observed in collapsars.

Fig.~\ref{fig:bns}(c-f) present the integrated GW spectra at different times after the merger. During the first $ \sim 30\,{\rm ms} $, the spiral arm generates strong GW emission owing to a significant deviation from axisymmetry. The contribution from different radii results in a roughly flat spectrum. As the disk starts to be accreted and become more axisymmetric, its GW emission decays with time. At $ t \gtrsim 100\,\ms $, the spiral arm vanishes, and density waves at $ R_d \approx 25\,\km $ govern the GW emission. The small radius of post-merger accretion disks results in a characteristic GW frequency of $ \fgw \approx 1.5\,{\rm kHz} $ [right panels, see Eq.~\eqref{eq:fGW}].

The SNR indicated in Tab.~\ref{tab:models} is likely too low for detection, even if considering the upper limit of BNS merger rate of\footnote{Post-merger massive disks with $ M_d \approx 0.1\,\msun $ power lGRBs \citep{Gottlieb2023e}. If standard short GRBs are more abundant in the Universe than lGRBs from mergers, then our simulated setup is uncommon among mergers, rendering this rate over-optimistic.} $ {\cal{R}_{\rm BNS}} \sim 10-1700\,{\rm Gpc\,yr^{-3}} $ \citep{Abbott2023} by future-generation GW detectors. Additionally, the GWs from the disk might be overshadowed by the GWs from the BH ringdown during the first several ms after the merger. This is particularly true considering unequal mass binaries tend to feature more massive disks for promising GW detectability, but also longer ringdown timescales. If the post-merger remnant is a meta-stable NS that peaks at $ \fgw \gtrsim 1\tilde{m}\,{\rm kHz} $ \citep{Shibata2006,Baiotti2008,Lehner2016}, then the ringdown can even last dozens of ms \citep{Ecker2024}.

\section{Comparison of Green's function }\label{sec:comparison}

\begin{figure}
    \centering
        {\includegraphics[width=7in]{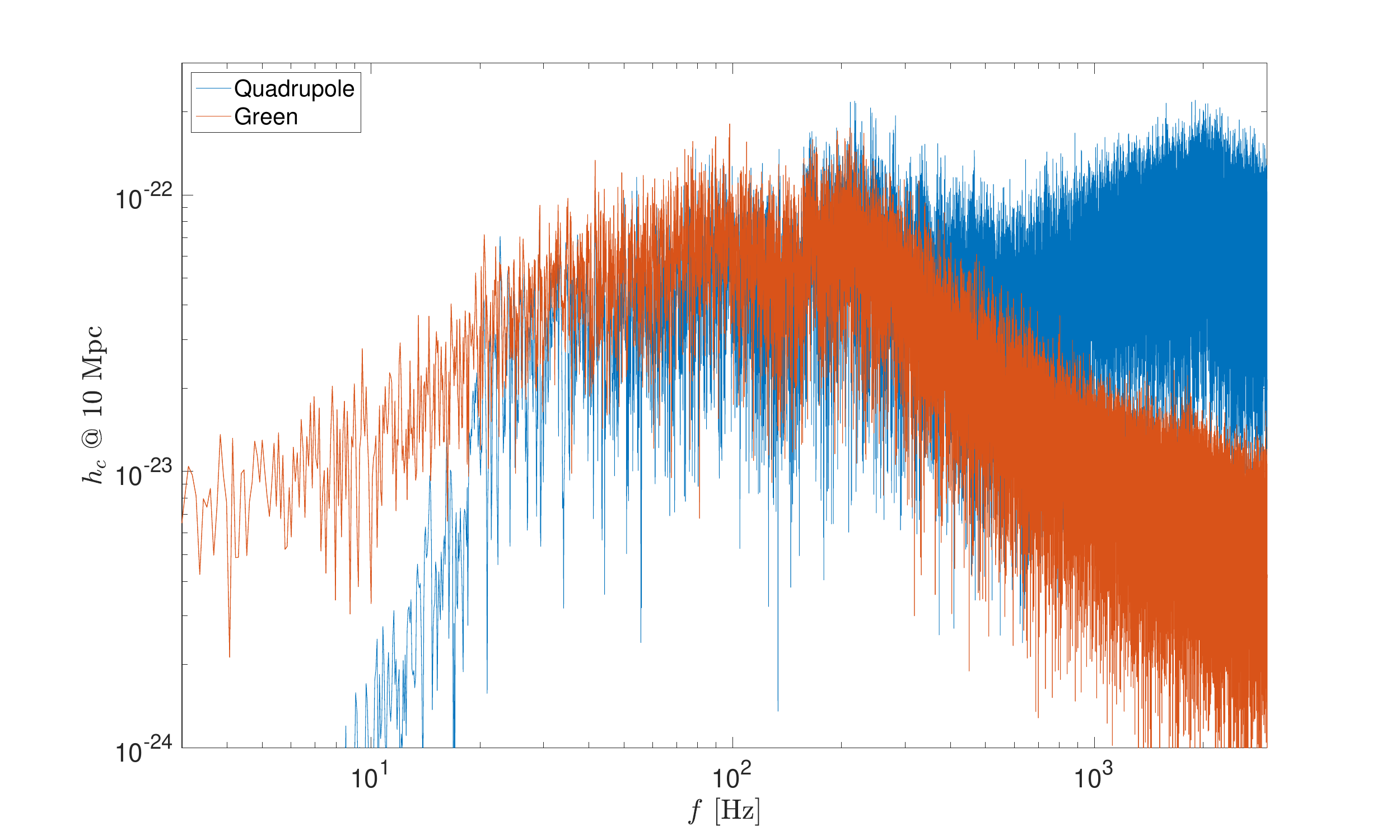}}
        \caption{The spectra in model $ C $ as calculated using the quadrupole approximation (blue) and the Green's function (red) for face-on observers.}
     \label{fig:comparison}
    \end{figure}

Figure~\ref{fig:comparison} shows a comparison between the Green's function calculation and the quadrupole approximation using the spectrum of model $C$. Around the peak frequency, $ \fgw \approx 100\,{\rm Hz} $, the two methods coincide. However, at low frequencies, the Green's function overestimates the spectrum, whereas, at high frequencies, the quadrupole approximation becomes invalid, resulting in a second peak at $ f > {\rm kHz} $. For the spectra shown in Fig.~\ref{fig:collapsar}, we use the minimum of the two methods.

\end{document}